%% file: digital-parks-nature.tex
\documentclass[fleqn,10pt]{wlscirep}

\usepackage{CJKutf8}

\usepackage{tabularx}
\usepackage[nolist]{acronym}
\usepackage[prependcaption]{todonotes} % not in white list
\usepackage{xargs} % not in white list
\usepackage{xspace}
\usepackage{multirow}
\usepackage{nicefrac}
\usepackage{float}
\usepackage{bbding}
\usepackage{bibunits}
\usepackage{paralist}

\usepackage{soul}
\usepackage[dvipsnames]{xcolor}

\input{acronyms}

\newcommand{\eg}{e.g.,\xspace}
\newcommand{\ie}{i.e.,\xspace}

\newcommandx{\sanja}[1]{\todo[linecolor=green,backgroundcolor=green!25,bordercolor=green,inline]{Sanja: #1}}
\newcommandx{\linus}[1]{\todo[linecolor=orange,backgroundcolor=orange,bordercolor=black,inline]{Linus: #1}}
\newcommandx{\andre}[1]{\todo[linecolor=teal,backgroundcolor=blue!25,bordercolor=black,inline]{Andre: #1}}
\newcommandx{\adam}[1]{\todo[linecolor=green,backgroundcolor=red!25,bordercolor=green,inline]{Adam: #1}}

\newcommand{\rev}[1]{\textcolor{black}{#1}}

\newcommand{\newrev}[1]{\textcolor{black}{#1}}

\newbool{showComments}
\booltrue{showComments}
\ifbool{showComments}{%
\newcommand{\fromA}[1]{\sethlcolor{pink}\hl{[A: #1]}}
\newcommand{\fromS}[1]{\sethlcolor{green}\hl{[S: #1]}}
\newcommand{\fromL}[1]{\sethlcolor{orange}\hl{[L: #1]}}
\newcommand{\fromGPT}[1]{\sethlcolor{yellow}\hl{[ChatGTP: #1]}}
}{
\newcommand{\fromA}[1]{}
\newcommand{\fromS}[1]{}
\newcommand{\fromL}[1]{}
\newcommand{\fromGPT}[1]{}
}

\title{Mobile Application Traffic Reveals Multifunctional Use Patterns in Parisian Parks}

\author[1,*]{André Felipe Zanella}
\author[2]{Linus W. Dietz}
\author[3,4]{Sanja Šćepanović}
\author[3,5]{Ke Zhou}
\author[6]{Zbigniew Smoreda}
\author[2]{Daniele Quercia}
\affil[1]{Telefónica Innovación Digital, Barcelona, Spain}
\affil[2]{King's College London, London, UK}
\affil[4]{The Uehiro Oxford Institute, University of Oxford, Oxford, UK}
\affil[5]{University of Nottingham, Nottingham, UK}
\affil[6]{Orange Research, Châtillon, France}
\affil[*]{andrefelipe.zanella@telefonica.com}

\begin{abstract}
    Urban parks play a key role in supporting public health. Landscape architecture typically considers parks through the lens of form and function. While past research on equitable access has focused mainly on park form, studies addressing functional uses have been constrained by limited scale and coarse measurement techniques. Existing efforts have partially quantified park functions through small-scale surveys and movement data (\eg GPS, check-ins) or general usage data (\eg CDR), but have not effectively captured the specific activities and motivations underlying park visits. As a result, our understanding of the functional roles urban parks play remains incomplete.
    To address this gap, we introduce a novel method that refines mobile base station coverage using antenna azimuths, enabling more precise distinction of mobile traffic within parks versus surrounding areas. Using Paris as a case study, we analyze a large-scale dataset of passively collected per-app mobile network traffic amounting to 492 million hourly records, representing a 35\% market share spanning 45 urban parks.
    We test two hypotheses: the Central-City hypothesis, which posits that multifunctional parks emerge in dense, high-rent areas due to land scarcity; and the Socio-Spatial hypothesis, which views parks as reflections of neighborhood routines and preferences. Our analysis shows that parks have distinctive mobile traffic signatures, differing from both their urban surroundings and from each other.
    By clustering parks based on temporal and application usage patterns, we identify three distinct functional types, Lunchbreak, Cultural, and Recreational parks, and analyze the traffic usage toward different motivations for visitation, i.e., pathways to realize parks' health-promoting potential. Centrally located parks (cultural and lunchbreak) display more diverse app usage and pronounced temporal variation, while suburban (recreational) parks reflect the digital behaviors of nearby communities, with app preferences aligned to neighborhood income.
    These findings demonstrate the value of mobile traffic as a proxy for studying the diversity of usage and activities within urban green spaces, with implications for park planning, public health, and well-being.
\end{abstract}

\begin{document}

\renewcommand\sectionautorefname{Section}
\renewcommand\subsectionautorefname{Section}
\renewcommand\subsubsectionautorefname{Section}
\renewcommand\figureautorefname{Figure}
\renewcommand\tableautorefname{Table}

\flushbottom
\maketitle

%%%%%%%%%%%%%%%%%%%%%%%%%%%%%%%%
\section{Introduction}
\label{sec:intro}
%%%%%%%%%%%%%%%%%%%%%%%%%%%%%%%%
\input{sections/1-introduction}

%%%%%%%%%%%%%%%%%%%%%%%%%%%%%%%%
\section{Results}
\label{sec:results}
%%%%%%%%%%%%%%%%%%%%%%%%%%%%%%%%
\input{sections/2-results}
\section{Discussion}
\label{sec:discussion}
%%%%%%%%%%%%%%%%%%%%%%%%%%%%%%%%
\input{sections/3-discussion}

%%%%%%%%%%%%%%%%%%%%%%%%%%%%%%%%
\section{Methods}
\label{sec:methods}
%%%%%%%%%%%%%%%%%%%%%%%%%%%%%%%%
\input{sections/4-methods}

%%%%%%%%%%%%%%%%%%%%%%%%%%%%%%%%

\bibliography{main}

%%%%%%%%%%%%%%%%%%%%%%%%%%%%%%%%

%%%%%%%%%%%%%%%%%%%%%%%%%%%%%%%%

\section*{Code Availability}
The Python code to calculate the park coverage scores, as well as replication of tests based on the RSCA and weekday-to-weekend traffic ratios are available at
\url{https://github.com/afzanella/digital_life_parks_replication}.

\section*{Author contributions statement}
A.F.Z., L.D. and S.S. conceived and conducted the experiments and co-drafted the manuscript. 
K.Z., Z.S. and D.Q. conceived experiments and edited the manuscript. 

\section*{Funding Declaration}
L.D. acknowledges support from the European Union’s Horizon 2020 research and innovation program under grant agreement no. 869764 (GoGreenRoutes).

\section*{Ethics Declarations}
\subsection*{Competing Interests}
The authors declare no competing interests.

\subsection*{Data Collection and Privacy:} 
The data obtained from this collection was collected by the operator for network management and research purposes, and temporarily stored and processed within a secure platform at their own premises, in full compliance with Article 89 of the General Data Protection Regulation (GDPR)~\cite{gdpr}. 
The data collection and processing were approved by the Data Protection Officer of the operator.
The researchers involved in this study only had access to such aggregates, whose spatiotemporal resolution ensures that no data subject can be re-identified from the data, which does not constitute personal data in terms of GDPR.

%%%%%%%%%%%%%%%%%%%%%%%%%%%%%%%%

%%%%%%%%%%%%%%%%%%%%%%%%%%%%%%%%
\input{sections/5-appendix}
%%%%%%%%%%%%%%%%%%%%%%%%%%%%%%%%

\end{document}

%% file: acronyms.tex
\acrodef{EPS}{Evolved Packet System}
\acrodef{TTI}{Transmission Time Interval}
\acrodef{UTLA}{Upper Tier Local Authority}
\acrodef{LAD}{Local Authority District}
\acrodef{CA}{Carrier Aggregation}
\acrodef{BSC}{Base Station Controller}
\acrodef{EPC}{Evolved Packet Core}
\acrodef{PDP}{Packet Data Protocol}
\acrodef{GMM}{GPRS Mobility Management}
\acrodef{GGSN}{Gateway GPRS Support Node}
\acrodef{NAS}{Non-Access Stratum}
\acrodef{RAN}{Radio Access Network}
\acrodef{GPRS}{General Packet Radio System}
\acrodef{UTRAN}{UMTS Terrestrial Radio Access Network}
\acrodef{SGSN}{Serving GPRS Support Node}
\acrodef{CN}{Core Network}
\acrodef{RNC}{Radio Network Controller}
\acrodef{MSC}{Mobile Switching Center}
\acrodef{MME}{Mobility Management Entity}
\acrodef{UMTS}{Universal Mobile Telecommunication Systems}
\acrodef{xDR}{eXtended Detail Record}
\acrodef{TAC}{Type Allocation Code}
\acrodef{PoP}{Point of Presence}
\acrodef{IMEI}{International Mobile Equipment Identity}
\acrodef{EMSISDN}{Encrypted Mobile Station International Subscriber Directory Number}
\acrodef{MSISDN}{Mobile Station International Subscriber Directory Number}
\acrodef{MME}{Mobility Management Entity}
\acrodef{VoLTE}{Voice over \ac{LTE}}
\acrodef{MVNO}{Mobile Virtual Network Operators}
\acrodef{CDR}{Call Detail Record}
\acrodef{IMSI}{International Mobile Subscriber Identity}
\acrodef{RAT}{Radio Access Technology}
\acrodef{LPWA}{Low Power Wide Area}
\acrodef{VMNO}{Visited Mobile Network Operator}
\acrodef{HMNO}{Home Mobile Network Operator}
\acrodef{DCH}{Data Clearing House}
\acrodef{GSMA}{\ac{GSM} Association}
\acrodef{GSM}{Global System for Mobile communications}
\acrodef{TAP}{Transferred Account Procedure}
\acrodef{LTE-M}{\ac{LTE} Machine Type Communication}
\acrodef{MTC}{Machine Type Communications}
\acrodef{NB-IoT}{Narrow Band \ac{IoT}}
\acrodef{IOT}{Inter Operator Tariff}
\acrodef{IoT}{Internet of Things}
\acrodef{M2M}{Machine-to-Machine}
\acrodef{FNO}{Fixed Network Operator}
\acrodef{ISP}{Internet Service Providers} 
\acrodef{ASP}{Application Service Providers}
\acrodef{IPX}{IP Packet Exchange}
\acrodef{2G}{Second Generation}
\acrodef{3G}{Third Generation}
\acrodef{4G}{Fourth Generation}
\acrodef{5G}{Firth Generation}
\acrodef{ADB}{Android Debug Bridge}
\acrodef{ASCI}{Advertising Standards Council of India}
\acrodef{ASN}{Autonomous System Number}
\acrodef{CDN}{Content Delivery Network}
\acrodef{DL}{Downlink}
\acrodef{DNS}{Domain Name Service}
\acrodef{EaaS}{Experiment as a Service}
\acrodef{ECDF}{Empirical Cumulative Distribution Function}
\acrodef{HTTP}{Hyper Text Transfer Protocol}
\acrodef{ICMP}{Internet Control Message Protocol}
\acrodef{ISP}{Internet Service Provider}
\acrodef{IXP}{Internet Exchange Point}
\acrodef{LTE}{Long Term Evolution}
\acrodef{MSC}{Message Sequence Chart}
\acrodef{E2E}{End-to-end}
\acrodef{QCI}{QoS Class Identifier}
\acrodef{NDT}{Network Diagnostic Tool}
\acrodef{QoE}{Quality of Experience}
\acrodef{QoS}{Quality of Service}
\acrodef{OS}{Operating System}
\acrodef{APN}{Access Point Name}
\acrodef{RMBT}{RTR Multithreaded Broadband Test}
\acrodef{RTT}{Round-Trip Time}
\acrodef{SIM}{Subscriber Identity Module}
\acrodef{SIMs}{Subscriber Identity Modules}
\acrodef{SLA}{Service-Level Agreement}
\acrodef{TCP}{Transmission Control Protocol}
\acrodef{UDP}{User Datagram Protocol}
\acrodef{UE}{User Equipment}
\acrodef{UL}{Uplink}
\acrodef{NRA}{National Regulatory Authority}
\acrodef{EC}{European Commission}
\acrodef{SGW}{Serving Gateway}
\acrodef{PGW}{Packet Data Network Gateway}
\acrodef{GTP}{GPRS Tunneling Protocol}
\acrodef{MNO}{Mobile Network Operator}
\acrodef{EU}{European Union}
\acrodef{HR}{home-routed roaming}
\acrodef{LBO}{local breakout}
\acrodef{IHBO}{IPX hub breakout}
\acrodef{VoIP}{Voice over IP}
\acrodef{IPG}{Inter-Packet Gap}
\acrodef{KS}{Kolmogorov-Smirnov}
\acrodef{FQDN}{Fully Qualified Domain Name}
\acrodef{MNC}{Mobile Network Code}
\acrodef{MCC}{Mobile Country Code}
\acrodef{MCCMNC}{\ac{MCC}-\ac{MNC}}
\acrodef{SIP}{Session Initiation Protocol}
\acrodef{g}{radius of gyration, (or gyration for short)}
\acrodef{MBB}{Minimum Bounding Box}
\acrodef{BBD}{Minimum Bounding Box Diagonal}
\acrodef{KPI}{Key Performance Indicator}
\acrodef{e}{Entropy}
\acrodef{DL}{downlink}
\acrodef{FL}{Federated Learning}
\acrodef{RoG}{Radius of Gyration}
\acrodef{ML}{Machine Learning}
\acrodef{IID}{independent and identically distributed}
\acrodef{BS}{Base Station}
\acrodef{NSA}{Non-Stand Alone}
\acrodef{SA}{Stand Alone}
\acrodef{POI}{Point of Interest}
\acrodef{AMF}{Access and Mobility Management Function}
\acrodef{RSCA}{Relative Symmetric Comparative Advantage}
\acrodef{ARI}{Adjusted Rand Index}

%% file: sections/1-introduction.tex
%\fromA{All requested changes have been implemented (see .md file). Main pending part is reviewing new content and streamlining new text from intro and discussion sections.}

%
Cities are shaped by the interaction between land use, movement, and daily routines. In aging societies, parks are prime health assets due to their multifunctionality: in dense and expensive urban areas, they take on many roles because space is limited and competition is high~\cite{fujita1996economics,glaeser2011triumph}, \rev{playing a key role in public health~\cite{Vlahov2007Urban, Wu2023Improved}, especially when exposure to at least 120 minutes per week in nature can be associated with improvements in well-being \cite{white2019spending}}. 
In contrast, parks also reflect the social traits, habits, and resources of nearby communities, forming neighborhood identities, seeing space as something created through everyday actions, cultural norms, and shared meanings~\cite{lefebvre2014production,low2016spatializing}.
The contrast between these two ideas allows for two competing premises to exist: the first is related to the \textit{Central-City} hypothesis, a top-down approach where park multifunctionality is defined by planners from the start to enrich the scarce space. The second is the \textit{Socio-Spatial} hypothesis, where parks are designed from a bottom-up approach, with their functionalities being shaped by the way people interact with them. 

% The demand for public spaces where people can exercise, relax, meet others, and enjoy culture is growing steadily. Urban parks serve as important places for such activities and play a key role in the general public health~\cite{Vlahov2007Urban, Wu2023Improved}. 
%
% Green space exposure is associated with reduced all-cause mortality, improved cardiovascular health~\cite{richardson2013role,muller2020effectiveness}, blood pressure reduction~\cite{shanahan2016health}, hypertension~\cite{aliyas2019built} and diabetes prevention~\cite{ponjoan2022impact}, and substantial mental health benefits including stress reduction~\cite{schram2018indicators,chen2021does}, mood improvement~\cite{olszewska2022therapeutic,zeng2023psychophysiological} and depression prevention~\cite{wang2019depressive,chen2021does}. 
%%Beyond individual physiology, parks serve as venues for individuals to stay active, to foster social cohesion in communities~\cite{jennings2019relationship}, and promote health equity~\cite{zhang2022assessing}.
% However, understanding how park design shapes use across different groups, and thus who can access these health benefits and what people do in these spaces, remains a major challenge, especially at large scale.

Understanding how people use these spaces differently across city limits—within both \textit{Central-City} and \textit{Socio-Spatial} hypotheses—is essential for urban planners who seek to improve park design, management, use and impact.
This often involves examining both the parks’ form---such as the amount and type of greenery or the presence of lakes---and their function, like encouraging sports, social interaction, or enjoying nature~\cite{ARRIBASBEL2022, dietz2024examininginequalityparkquality}. Though form and function are linked, prior studies have focused on form, treating parks mainly as green infrastructure~\cite{semeraro2021planning,li2015comparison}. Others have studied parks’ functional benefits, especially health gains and, particularly, health outcomes such as physical activity~\cite{takano2002urban,cox2017doses}.
\rev{Urban studies have also grouped parks into categories. Researchers have proposed typologies for specific cities~\cite{zhou2024visitation,cao2021functional,Liu2020Categorization,ibes2015multi}, shaped by the goals of each analysis.}
% To better understand how visitors use parks in the digital age and to guide planning across the spatial diversity of cities, we analyze detailed patterns of mobile phone traffic in individual parks for the first time. This novel approach enables us to characterize park functions at scale while capturing the spatial and socioeconomic heterogeneity shaping visitor behaviors.

The growing availability of mobile phones offers new opportunities to study urban space use through the data they passively generate, revealing detailed information about how urban spaces are used \cite{ghahramani2020urban,vscepanovic2015mobile}, reliably showing where and when people are present and how they move in specific areas \cite{dong2024defining}. Early studies used call detail records (CDRs) from mobile networks to distinguish park zones from other types of land use \cite{toole2012inferring}, while recent work has used GPS data to study park accessibility \cite{mears2021mapping, filazzola2022using}, the impact of the COVID-19 pandemic \cite{kim2023mobile, Zhao2023Change}, and inequalities in park access based on space, socioeconomic status, and activity~\cite{xiao2019exploring, ren2022evaluating,guo2019accessibility}.
Overall, smartphone data has so far been useful for identifying where and when people visit parks, but less helpful for explaining why they visit or what they do there \cite{Heikinheimo2020Understanding}.

\rev{This limitation in understanding \textit{what} users might be doing in parks comes from the data sources. GPS \cite{liu2021application, zhai2018spatial, Zhao2023Change} and CDR \cite{zhai2018using} are limited to showing visitors' mobility and journey to parks. App-specific traffic from single platforms (\eg Strava \cite{Heikinheimo2020Understanding}, Flickr \cite{Song2020Using}), geotagged images \cite{Song2020Using,dietz2024exploratory} and social media posts \cite{Ghahramani2021Tales} limit the observed behavior to a single activity and may suffer from limited representativeness.
With the densification of mobile networks and rise in adoption of smartphones, app-level mobile traffic can be an alternative source to help understand spatiotemporal phenomena at country \cite{Zanella2022Impact} and city \cite{zanella2024modeling} scale, and can be an interesting proxy to help understand \textit{what} visitors might be doing in parks. However, accurately localizing traffic at small urban areas remains an open challenge, often relying on square grids or Voronoi cell overlays \cite{candia2008uncovering} to estimate smartphone utilization in park boundaries. 
}

\rev{To address these gaps, we develop a method to assign application-level mobile network traffic to fine-grained urban areas using data from a major telecom provider in Paris. This approach enables large-scale, passively collected observation of smartphone use within parks, providing new insight into how people use these spaces across time, location, and socioeconomic context.
Using this framework, we examine how smartphone activity differs between parks and other urban areas, classify parks based on patterns of mobile app use, and test how the \textit{Central-City Multifunctionality} and \textit{Socio-Spatial Differentiation} hypotheses coexist in practice.}
Our research addresses the following questions:
\begin{inparaenum}[\itshape i)\upshape]
    \item How can passively collected mobile app traffic be reliably assigned to urban parks?
    \item How does smartphone app use differ between parks and other city areas?
    \item How does smartphone use vary across parks, and what temporal, spatial, and behavioral patterns are visible?
    \item Can park use be classified based on mobile app usage, and what functional types of park use emerge?
    % \item How do park features and the socioeconomic profiles of nearby neighborhoods relate to mobile traffic within parks?
    \item Can mobile app usage reveal how the \rev{two spatial hypotheses of park use} coexist in urban environments, and how they relate to city space? 
\end{inparaenum}

%% file: sections/2-results.tex
\label{sec:traffic_patterns}
We use large-scale traffic data from a major mobile network operator in Paris, collected between January 31 and May 31, 2023 (see \autoref{subsec:data_collection} for data and ethical details).
First, we identified parks where mobile traffic could be reliably linked to park visitors.
\newrev{Thus, our study has a focus on medium to large urban parks (minimum 1.2 ha, median 25.5 ha), each with sufficient cellular antenna coverage to allow reliable attribution of traffic to the park area (see \autoref{subsec:park_selection_methodology} for methodology details).}
For the selected 45 parks, we examined usage patterns and found clear differences in mobile app use between parks and other parts of the city, along with distinct time-based patterns.
These usage profiles allowed us to group parks into three types based on app use, which we linked to the socioeconomic indicators of nearby neighborhoods, validating the coexistence of both the Central-City and Socio-Spatial hypotheses in Paris.

%------------------------
%------------------------
\subsection{Reliable Assignment of Traffic to Parks}
\label{subsec:park_selection}
%------------------------
%------------------------

\begin{figure}[t]
    \centering
    \includegraphics[width=0.95\linewidth]{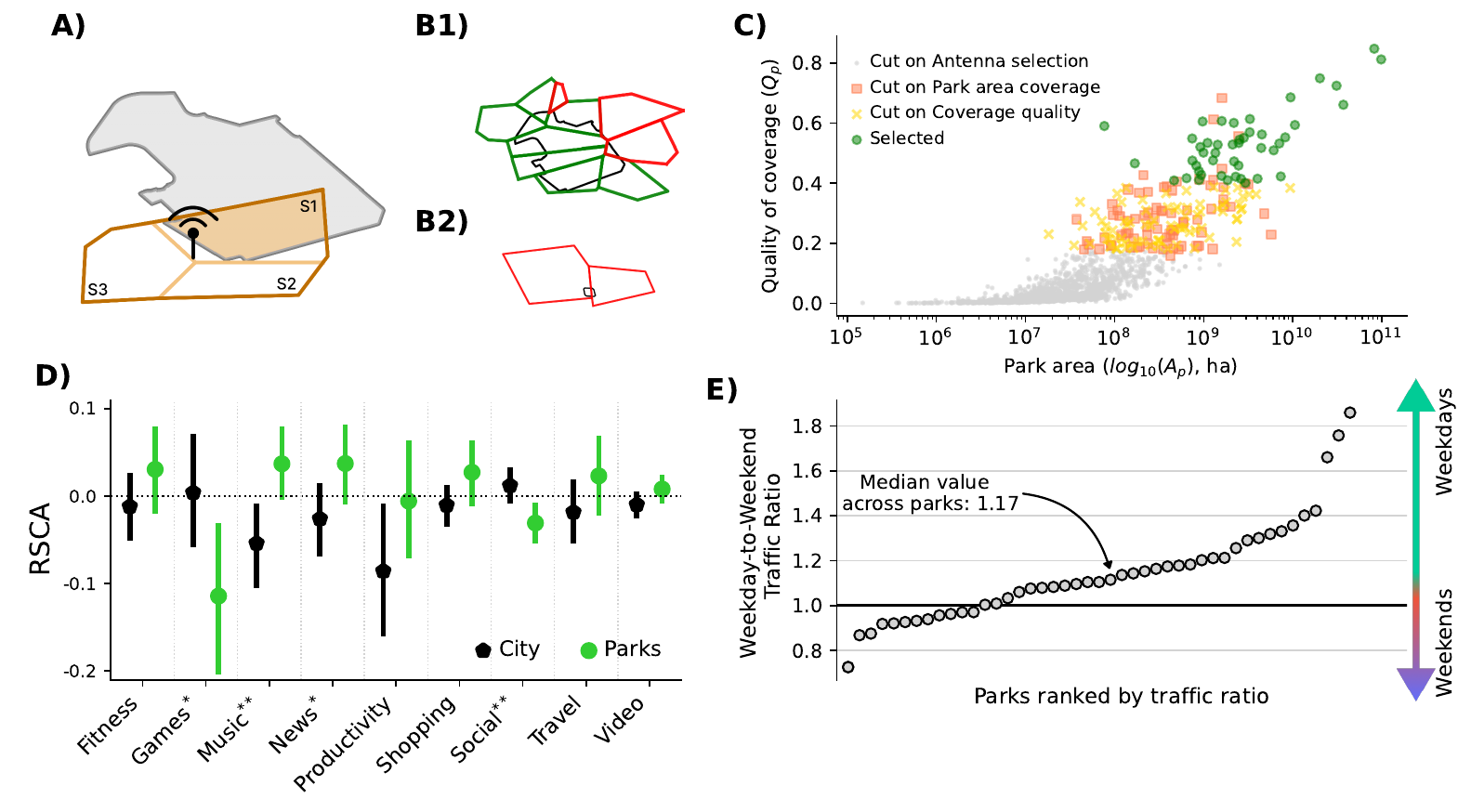}
    \caption{
    Top: Methodology for selecting parks that can be reliably analyzed.
    (A) Considering a base station, we leverage the coverage sectors to allow finer-grained Voronoi geometries, which allow a selection of geometries with better overlap with parks. 
    Example of (B1) good coverage (top - Jardin de Luxembourg) \emph{vs.} (B2) poor coverage (bottom - Square Samuel Rousseau);
    (C) Plotting park area (x-axis) against the Quality of Coverage $Q_{p}$ (cf. \autoref{subsec:park_selection_methodology}, y-axis). Larger parks generally have better coverage, leading to a focus on bigger parks.
    From 1,641 named parks in Paris, we selected 45 (min: 1.2 ha; median: 25.5 ha; max: 995 ha) for our study.
    Bottom: Traffic analysis across parks.
    (D) RSCA distribution in parks and the city, indicating app popularity (RSCA $>$ 0) or unpopularity (RSCA $<$ 0). Markers show means, lines indicate $95\%$ confidence intervals. $*$ and $**$ mark significant differences.
    (E) Ratio of traffic on weekdays \emph{vs.} weekends for individual parks (gray), park average (green), and city average (black). Values $>1$ indicate higher traffic on weekdays.}
    \label{fig:park_selection_methodology}
\end{figure}

%%%%%%%%%%%%%%%%%%%%%%%%%%%%%%%%
% Introducing the overlap between park and antenna geometries
%%%%%%%%%%%%%%%%%%%%%%%%%%%%%%%%
A central contribution of our research is moving beyond coarse-grained spatial analysis of mobile traffic, which traditionally relies on Voronoi geometries to approximate base station coverage. %\cite{aurenhammer1991voronoi}.
We refine earlier approaches that incorporated azimuth angles, \ie dividing base station coverage into antenna sectors to optimize coverage\cite{janecek2015cellular,zanella2024characterizing}, to isolate mobile traffic at specific small-scale urban spaces, such as parks.
Our method is illustrated in \autoref{fig:park_selection_methodology}.
In Jardin du Luxembourg (\autoref{fig:park_selection_methodology}A), the traditional Voronoi geometry (dark orange outline) covers the park but also a wide surrounding area.
By considering the base station's three antenna sectors (S1–S3), we can define a more precise coverage area (light orange outlines).
Sector S1 primarily covers the park (orange area) and is the most suitable for our analysis, while S2 and S3 point in other directions and should be excluded, as they do not cover the park. 

%%%%%%%%%%%%%%%%%%%%%%%%%%%%%%%%
% Park selection
%%%%%%%%%%%%%%%%%%%%%%%%%%%%%%%%
Concretely, we followed three steps:
1) selecting antennas that sufficiently cover parks (\autoref{fig:park_selection_methodology}A/B),
2) identifying parks served by those antennas (\autoref{fig:park_selection_methodology}B), and
3) retaining parks where the assigned traffic exceeds the thresholds defined for reliable attribution (\autoref{fig:park_selection_methodology}C).
This process results in a final set of 45 parks for analysis (further details of studied parks in Supplementary Information).
We observe a direct correlation between the Quality of Coverage metric and park area (\autoref{fig:park_selection_methodology}C), indicating that our sample mainly includes medium- to large-sized urban parks.
This outcome is expected, as larger parks are more likely to be served by dedicated antennas. Even though modern cellular networks are dense in urban areas, the individual coverage areas are often too small to cover small neighborhood parks.
\newrev{This data limitation reflects the current spatial granularity of mobile network deployment and prevented us from reliably analyzing small and pocket-sized urban parks.}

%------------------------
%------------------------
\subsection{The Uniqueness and Diversity of Smartphone use in Parks}
%------------------------
%------------------------

To compare mobile app usage in parks with other urban spaces, we selected the 41 most relevant mobile apps, which together account for almost 80\% of traffic (see \autoref{subsec:mobile_apps}). These apps were grouped into nine categories: Fitness, Games, Music, News, Productivity, Shopping, Social, Travel, and Video (see \autoref{tab:apps}, \autoref{subsec:mobile_apps}).

%%%%%%%%%%%%%%%%%%%%%%%%%%%%%%%%
% Traffic is unique in parks vs. the city
%%%%%%%%%%%%%%%%%%%%%%%%%%%%%%%%
% -----
\subsubsection*{People Tend to Use Different Mobile Apps in Parks Compared to the Rest of the City}
% -----
\autoref{fig:park_selection_methodology}D shows the \ac{RSCA} for each app category in parks and in other areas of the city, \rev{allowing us to perform an analysis of comparative advantage, \ie observe the over- or under-utilization of mobile applications in parks versus the city.}

\noindent \textit{Park-specific applications.}
Apps in the Music and News categories stand out for being used significantly more in parks than in the rest of the city. This difference in mean usage is statistically significant ($p < 0.005$ for Music and $p = 0.049$ for News (details in the Supplementary Information). Categories such as Fitness, Video, Shopping, and Travel also have higher average usage in parks, but with statistically non-significant differences.

\noindent \textit{City-specific applications.}
In contrast, Social and Games apps are used significantly more outside parks. The differences in their mean usage are statistically significant ($p = 0.005$ for Social, $p = 0.027$ for Games).
Social apps show negative RSCA values in parks and positive values elsewhere, suggesting under-use in park settings. Games follow a similar pattern: they are underused in parks but more evenly distributed across the city.

Despite these trends, RSCA confidence intervals in parks show substantial variation across categories (\eg Games), suggesting that app usage within parks is not uniform, requiring a more detailed park-level analysis.

%%%%%%%%%%%%%%%%%%%%%%%%%%%%%%%%
% Temporal traffic has variations across parks
%%%%%%%%%%%%%%%%%%%%%%%%%%%%%%%%
% -----
\subsubsection*{Temporal Patterns in Park Traffic Reflect Work–Leisure Utilization}
% -----
To understand temporal patterns in park usage, we divided traffic volumes into weekdays and weekends\cite{furno2016tale}, calculating the median daily volume for each. Approximately 75\% of parks experience higher traffic on weekdays than on weekends (\autoref{fig:park_selection_methodology}E). However, the median ratio of weekday to weekend traffic in parks is $1.17$, compared to a slightly higher ratio of $1.19$ in the rest of the city. This suggests that although mobile traffic generally declines on weekends, the decrease is less pronounced in parks.
The weekday-to-weekend traffic ratio varies considerably across parks, ranging from $0.72$ to $1.86$. To understand what types of parks are used more during different periods, we examined both ends of this spectrum.

\noindent{\textit{Parks with high weekend traffic.}} The three parks with the highest weekend traffic—Parc départemental de la Fosse Maussoin, Parc Georges Brassens, and Parc de la Vallée-aux-Loups—are primarily oriented toward nature-based activities. Parc départemental de la Fosse Maussoin and Parc de la Vallée-aux-Loups, both in the Parisian suburbs, feature dense green spaces. Parc Georges Brassens, located more centrally, includes a pond, meadows, tree groves, and several themed gardens, including a vineyard. These parks, which attract more visitors on weekends, are mainly situated in residential areas.

\noindent{\textit{Parks with high weekday traffic.}} Conversely, the three parks with the highest weekday traffic—Jardins Abbé Pierre, Parc Monceau, and Parc Suzanne Lenglen—are located in or near areas with high concentrations of offices, schools, or government buildings. Jardins Abbé Pierre and Parc Monceau are centrally located, near universities and business districts. Parc Suzanne Lenglen, on the southern edge of Paris near Issy-les-Moulineaux, includes a large multi-sport facility used for classes and is close to both the Ministry of Defense and a major digital business hub—factors that likely contribute to its high weekday traffic.

%------------------------
%------------------------
\subsection{Three Park Types Emerge from Traffic Patterns}
\label{subsec:parks_cluster}
%------------------------
%------------------------

%%%%%%%%%%%%%%%%%%%%%%%%%%%%%%%%
% App usage across clusters
%%%%%%%%%%%%%%%%%%%%%%%%%%%%%%%%
% -----
% \noindent \textbf{Clustering Traffic Patterns across Parks}
% -----
We used each park’s per-application RSCA values and its weekday-to-weekend traffic ratio as features in spectral clustering (see \autoref{sec:clustering}) to classify parks according to their smartphone usage patterns. Based on the Silhouette score and additional qualitative analysis, we selected $k=3$ as the final number of clusters (details for the clustering methodology in the Supplementary Information), as depicted in \autoref{fig:park_categories_insights}A.
Examining the parks within each cluster reveals notable differences in temporal traffic patterns (\autoref{fig:park_categories_insights}B) and mobile app category preferences (\autoref{fig:park_categories_insights}C). We also observe a spatial pattern in the distribution of clusters across the city. 

\subsubsection*{The Functional Park Types in Paris: Cultural, Lunchbreak and Recreational}
We identified three primary functional park profiles; to determine the names for the clusters and to understand how people use these parks, we analyzed Flickr image tags associated with each park to assign meaningful names that reflect the functional identities of clusters (details in the Supplementary Information). 
Tags with higher relative importance highlight thematic differences among clusters—such as natural and urban features for Lunchbreak parks, arts and festivals for Cultural parks, and lower-density, French-oriented tags for Recreational parks.
% This approach allowed us to assign meaningful names reflecting the functional identities of park clusters based on the distinct tag patterns of images shared within them (methodology details on \autoref{subsubsec:clusters_flickr}).

\begin{figure}[t]
    \centering
    \includegraphics[width=0.95\linewidth]{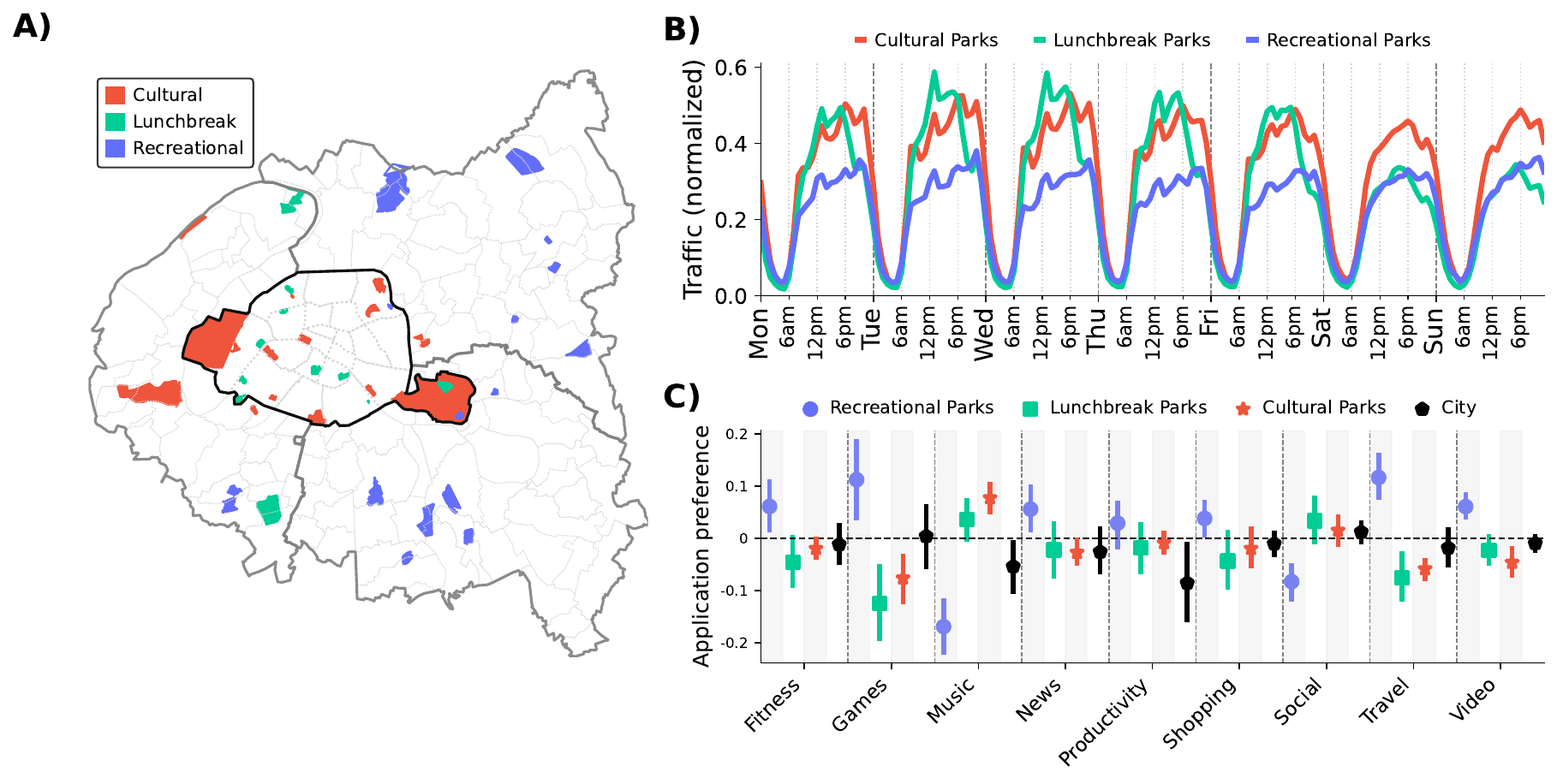}
    \caption{
    (A) Park types across Paris.
    (B) Median weekly traffic per cluster for parks within each cluster.
    (C) RSCA by application and park cluster. Points show mean values, while lines indicate $95\%$ confidence intervals across parks in each cluster.
    }
    \label{fig:park_categories_insights}
\end{figure}

\textit{Recreational parks} (in blue) are in the suburban areas of Paris, known as the \emph{Banlieue}. These areas include separate cities in the departments of Hauts-de-Seine, Seine-Saint-Denis, and Val-de-Marne. According to the 2017 census, 68\% of the people living in metropolitan Paris reside in these less densely populated, mostly residential zones. Examples of parks in this group include Parc départemental de la Fosse-Maussoin and Domaine départemental de la Vallée-aux-Loups. Parks in this group have the lowest traffic use compared with other parks and the rest of the city: 57\% less median traffic than the city's median. In contrast, the median traffic at parks in the other two clusters is close to the city's overall level: Lunchbreak parks have only 8\% less traffic, while Cultural parks have 3\% more.

\textit{Lunchbreak parks} are in densely built areas near business districts and universities. Their traffic patterns are distinct: traffic is much higher on weekdays, with median traffic 58\% greater than on weekends as shown in \autoref{fig:park_categories_insights}B. In \autoref{fig:park_selection_methodology}E, these parks appear on the right side of the plot.
Lunchbreak parks are smaller, with a median area just 36\% of that of other parks. Their weekday traffic patterns, closeness to business areas, and the types of apps used on weekdays  (breakdown of application preference between weekdays and weekends is the Supplementary     Information) suggest that office workers use these parks during work breaks.
Examples include the Jardin du Luxembourg and the Jardin des Plantes. While these parks may have cultural attractions, their location near dense work and study areas, along with their usage patterns, suggests they are predominantly used by workers or students taking breaks during the week.

\textit{Cultural parks} (in red) are known for their attractions, such as the Champ de Mars with the Eiffel Tower, Jardin des Tuileries, and Parc de la Villette, which includes a museum and concert venues.
These parks show the most even traffic distribution between weekdays and weekends, with weekdays having only 7\% more median traffic. Their steady popularity likely stems from their historical significance and location in a city with high tourist activity. This also explains why they have the highest median traffic among the three clusters, with 3\% more than the city's median.

To further validate the clusters, we conducted a qualitative analysis of several key parks.
The highest weekday-to-weekend traffic ratio is in Parc Monceau (Lunchbreak), located in a mixed-use area with many offices and embassies. It shows a sharp drop in traffic on weekends, despite being surrounded by residential buildings, suggesting that local residents may not see it as a major weekend destination, or that the area's population declines significantly outside the workweek.
Parks with above-average traffic are mostly classified as Cultural or Lunchbreak, with the highest overall traffic seen in Jardin des Plantes, located in central Paris near Sorbonne University.
The parks with the lowest traffic are those with a high density of natural features and are classified mainly as Recreational. We suggest that the nature-focused design of these parks may influence visitor behavior, leading to reduced smartphone use.
Parc Pierre-Lagravère and Parc Jean-Moulin–Les Guilands were also classified as Cultural, despite being in suburban Paris. Both underwent recent renovations that introduced modern buildings and wide open spaces---features that differ from other suburban parks, which might have led to the change in smartphone patterns.

%%%%%%%%%%%%%%%%%%%%%%%%%%%%%%%%
% App usage across clusters
%%%%%%%%%%%%%%%%%%%%%%%%%%%%%%%%
%------------------------
%------------------------
\subsubsection*{App Usage Varies Among Park Types}
%------------------------
%------------------------
We next explore how mobile application preferences change across park types (\autoref{fig:park_categories_insights}C), conducting a series of experiments to validate dissimilarities (detailed in the Supplementary Information).
\textit{Cultural} parks show lower usage across most app categories, suggesting that visitors use a limited range of apps. Social media and music streaming apps stand out with above-average usage, likely reflecting tourists who focus on social engagement rather than routine activities such as shopping, gaming, productivity, or news while exploring the city.
\textit{Lunchbreak} parks show a similar pattern, with most app categories displaying neutral or slightly negative \ac{RSCA} values, except for social and music apps. This suggests that both centrally located clusters (Cultural and Lunchbreak) similarly influence smartphone utilization, a pattern also seen around the general areas of Paris (shown in black in \autoref{fig:park_categories_insights}C).
Overall, Cultural and Lunchbreak parks show similar (statistically non-significant) differences in app preferences compared to the rest of the city, with the exceptions of music apps and video apps (further details in Supplementary Information).

The \textit{Recreational} parks present distinct app preference patterns not only from the rest of the city, but also from the other two park clusters, with significant differences for almost all app categories (statistical tests detailed in Supplementary Information).
Recreational parks show over-utilization across most app categories, except for music and social media. Notably, they are the only parks with increased use of fitness apps, along with higher engagement in news and travel categories.
Combined with the insight that \textit{Recreational} parks have the lowest overall traffic (\autoref{fig:park_categories_insights}B), these findings suggest that these parks serve a unique role, offering a setting for relaxation and everyday routines while encouraging physical activity and limiting social media use. When visitors use their phones, they tend to favor apps for reading and planning.

Beyond their technical classification, these three functional park types reflect distinct health-relevant usage patterns with important equity implications. 
Lunchbreak parks may have utilization patterns linked to occupational populations with regular daytime proximity to central districts. 
\rev{Access to outdoor green spaces outside of workplaces can have an effect in reducing stress \cite{lottrup2013workplace}, offering a more pleasant place to walk during breaks and release tension \cite{de2017effects}, even when they remain connected to their phones.}
%
% Cultural parks, displaying high temporal diversity and engagement with Social and Music applications, indicate potential for social cohesion and community engagement. 
%
Recreational parks in suburban areas deserve closer study for their \newrev{potential} impact on health equity. 
\newrev{Lower overall smartphone use and under-use of social applications in Recreational parks suggest that visitors might engage in fewer passive digital behaviors. Although this study does not directly measure health outcomes, previous works have linked them to lower screen time and potential well-being benefits~\cite{pieh2025smartphone}. This leaves space for future works to further study the under-explored topic of the joint effects of screen time and green time on health and well-being~\cite{oswald2020psychological}, especially in the context of urban parks.}
% Lower overall smartphone use and underuse of social applications suggest that these parks may help reduce screen time, in contrast to the patterns seen in the other two clusters.

%------------------------
%------------------------
\subsection{Two Lenses on Park Multifunctionality: Economic Pressure versus Cultural Mediation}
\label{sec:clusters_socioeconom}
%------------------------
%------------------------

\begin{figure}[t]
    \centering
    \includegraphics[width=0.8\linewidth]{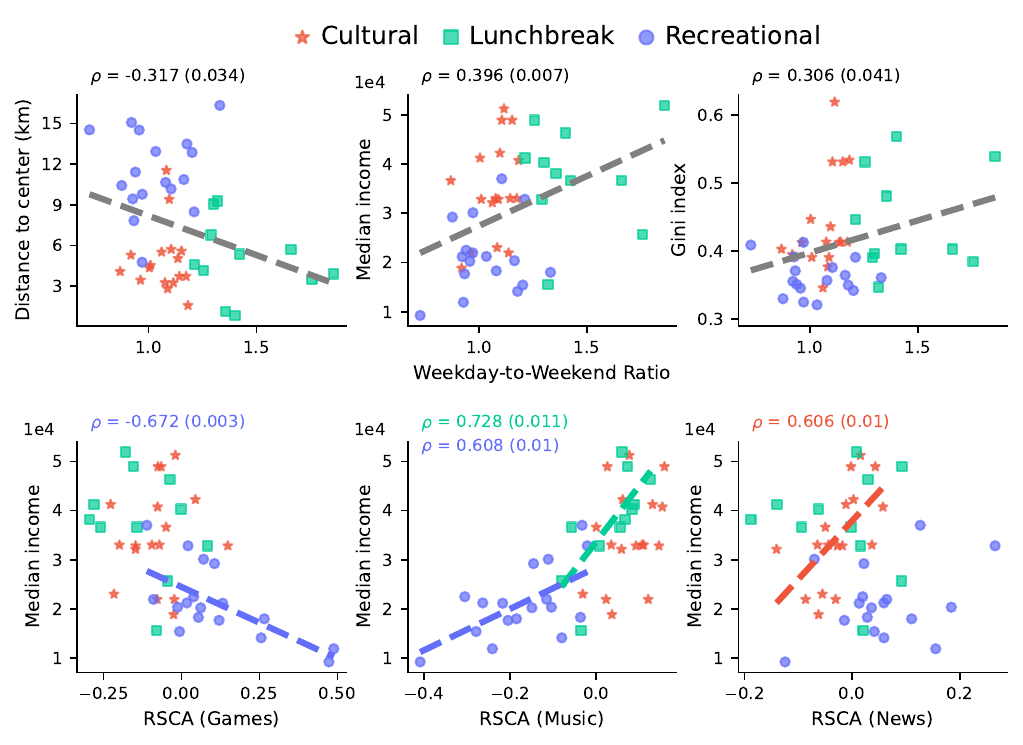}
    \caption{
    Relation between weekday to weekend traffic ratio (top) and RSCA (bottom) versus socioeconomic indicators, for park categories. Pearson correlation coefficients $\rho$ and $p$-values (in parentheses) are indicated on top. 
    }
    \label{fig:scatter_socioeconomics}
\end{figure}

\rev{To investigate whether socioeconomic factors underlie the observed smartphone 
patterns, we evaluate two competing hypotheses using Pearson correlations 
between traffic consumption, application preferences, and socioeconomic 
indicators.
Socioeconomic indicators are collected at the IRIS level, the smallest 
geographical unit in the French Census~\cite{census2017}.
A park is linked to an IRIS, if the IRIS contains the majority of each park's area.
As our data is aggregated at the antenna level, we cannot infer 
visitors' home locations or distinguish residents from non-residents. We therefore treat socioeconomic indicators as attributes of each park's surrounding context. This is supported by evidence of strong socioeconomic segregation in large cities~\cite{nilforoshan2023human}, which limits cross-neighborhood mobility and strengthens the correspondence between an area's socioeconomic profile and that of its park visitors.}

% --
\subsubsection*{The Central-City Hypothesis}
% --

The Central-City hypothesis describes a top-down approach: it posits that centrally located parks exhibit greater diversity in human activity due to the combined effects of agglomeration and high land values. In major cities like Paris, space in dense and expensive areas must serve multiple functions to justify its use---office workers, students, cafe-goers, and transit users.

\noindent \textit{Park location and weekday usage.}
\autoref{fig:scatter_socioeconomics} (top) presents the relationship between the weekday-to-weekend traffic ratio and three key socioeconomic indicators: distance from the city center, median household income, and economic inequality (Gini Index).
We observe a modest negative Pearson correlation between distance to the city center and the weekday-to-weekend app traffic ratio ($\rho = -0.317, p = 0.034$). Parks closer to the center exhibit higher weekday activity, while those farther away tend to attract more visitors on weekends. This pattern aligns with the expectation that centrally located spaces, such as Lunchbreak parks, cater to weekday populations. While the weekday-to-weekend ratio of Cultural parks is lower, their overall traffic demands during weekdays are still on par with Lunchbreak parks (evidenced by \autoref{fig:park_categories_insights}B). 

\noindent \textit{Affluence and multifunctionality.}

\newrev{Median income in the surrounding area shows a positive correlation with the weekday-to-weekend ratio ($\rho = 0.396, p = 0.007$). This correlation reflects not income as a direct driver, but rather the co-location of high income with urban centrality in Paris, as observed in \autoref{fig:scatter_socioeconomics} through the distribution of park clusters and median income levels. In dense, affluent central districts, high land values create acute economic pressure: the scarcity and cost of space force parks to serve multiple constituencies simultaneously (office workers, students, commuters, and tourists) to justify their continued use and upkeep. This multi-functionality is evident in the distinct weekday-to-weekend signatures of the two central park types: Lunchbreak parks are located precisely in office and university concentrations and show the steepest weekday-to-weekend traffic ratios (up to $1.58$), while Cultural parks show more balanced weekday-to-weekend usage.}

\newrev{This differentiation reveals that the weekday-to-weekend pattern can be explained by functional pressure and location, not income per se. Cultural and Lunchbreak parks are indeed predominantly situated in higher-income neighborhoods, whereas Recreational parks are located in areas with low to average incomes; similarly, the Gini index pattern observed in \autoref{fig:scatter_socioeconomics} ($\rho = 0.306, p = 0.041$) mirrors this spatial distribution, with Lunchbreak and Cultural parks more common in areas of greater inequality (Gini $> 0.5$) and Recreational parks in areas of lower inequality (Gini $< 0.4$). Together, these patterns support the Central-City hypothesis: centrality and economic scarcity drive parks to become multifunctional support spaces, providing affordable, accessible opportunities to decompress in areas where alternative leisure options are costly.
These spaces serve either as break spaces for work/study life (Lunchbreak) or as cultural and tourist destinations (Cultural).}

% Median income in the surrounding area shows a positive correlation with the weekday-to-weekend ratio ($\rho = 0.396, p = 0.007$). Cultural and Lunchbreak parks are predominantly situated in higher-income neighborhoods, whereas Recreational parks are located in areas with low to average incomes. The Gini index pattern mirrors this result ($\rho = 0.306, p = 0.041$): Lunchbreak and cultural parks are more common in areas with greater inequality (Gini $>0.5$), while Recreational parks are found in areas with lower inequality (Gini $<0.4$).

% These correlations support the Central-City hypothesis: centrally located, premium-area parks exhibit concentrated weekday use and multifunctionality. Lunchbreak parks in particular show reduced weekend demand, likely due to their role in serving surrounding office and academic populations. 
%
% These urban parks function as crucial support spaces—providing affordable opportunities to decompress in areas where leisure options are costly---illustrating the economic pressures that drive multifunctional use in high-rent districts, either as escape spaces for the work/study life (Lunchbreak) or from tourism, sightseeing and cultural activities in Paris (Cultural).

% --
\subsubsection*{The Socio-Spatial Hypothesis}
% --

The Socio-Spatial hypothesis describes a bottom-up process: park use patterns reflect the habits, resources, and cultural practices of surrounding communities~\cite{lefebvre2014production,low2016spatializing}. In this view, suburban park multifunctionality emerges from social mediation and lifestyle norms rather than from urban form pressures.

\noindent \textit{Application preferences and income levels.}
We examine the relationship between application category preferences (measured via \ac{RSCA} values) and socioeconomic indicators (\autoref{fig:scatter_socioeconomics}, bottom). 
\newrev{The socioeconomic context of the areas surrounding parks may not fully reflect the characteristics of park visitors, since individuals can travel from other parts of the city to visit them. Because the data used in this study does not distinguish between nearby residents and cross-town visitors, the IRIS-based socioeconomic indicators should be interpreted as contextual proxies rather than direct visitor attributes. This is especially relevant in Paris, where the dense public transportation networks enable relatively easy cross-neighborhood access.
}
\newrev{However, previous works have observed that travel distance can limit park visitation~\cite{tu2020travel,dietz2024exploratory} and that, even in well-connected cities, large-scale socioeconomic segregation structures daily movement patterns: people disproportionately encounter and visit places with socioeconomic profiles similar to their own~\cite{nilforoshan2023human,moro2021mobility}. This suggests that, while the data used cannot distinguish between locals and visitors, transit accessibility may not fully erase the socioeconomic structuring of exposure and visitation contexts.}

\newrev{We observe the following statistically significant relations between mobile application preferences and income levels from the regions where parks are located:}

\noindent \textit{Gaming apps:} In Recreational parks, gaming usage declines with increased median income, suggesting that visitors \newrev{of parks in} lower-income areas may engage more in passive entertainment activities. Cultural and Lunchbreak parks show lower gaming usage overall, independent of income.

\noindent \textit{Music apps:} Music app usage increases with income in both Lunchbreak and Recreational parks; Cultural parks display a similar but non-significant trend.

\noindent \textit{News apps:} Only Cultural parks show a significant positive correlation between income and news app usage, indicating greater consumption of news content in higher-income contexts.

These findings support the Socio-Spatial hypothesis: suburban Recreational parks' \rev{app signatures are systematically associated with the socioeconomic context in which the parks are embedded}. In lower-income neighborhoods, parks serve as low-cost leisure venues \newrev{for the visitors they receive (whether locals or outsiders)}, with higher engagement in gaming.
In wealthier suburban contexts, Recreational parks see greater use for music and potentially productivity-related activities. For example, the News and Information category in Recreational parks shows a positive (though non-significant) correlation with income ($\rho = 0.317, p = 0.215$). This aligns with the Socio-Spatial view that park activities are embedded in local digital cultures and lifestyles.

%% file: sections/3-discussion.tex
%
Using the proposed framework, we are able to accurately isolate mobile traffic within a defined set of parks.
This enables the first large-scale analysis of how \rev{weekly aggregations of} mobile traffic and applications are consumed across parks. We identify three main functional park types across a set of 45 parks in Metropolitan Paris, each showing distinct smartphone patterns that reflect different user behaviors and activities.
While our three park types partially align with existing categorizations~\cite{KONG2022104482,ZHANG201827}, our work establishes a new, data-driven classification grounded in behavioral usage data, exploring how competing hypotheses for spatial utilization of parks coexist in the city of Paris.

%------------------------
%------------------------
\subsection{Main Findings}
%------------------------
%------------------------

% ---
% Results from the plot of Fig 1, about the RSCA in park vs. city and the wd/we ratio
\subsubsection*{Smartphone Usage is Diverse in Parks and Different From the Rest of the City}
% ---
Our findings show that mobile application usage via smartphones differs significantly in parks compared to other parts of the city. This supports previous land-use studies that identified green spaces as a distinct category in mobile phone activity~\cite{toole2012inferring,furno2016tale}, and further reveals that fitness, music, news, shopping, and travel apps are more frequently used in parks, while social media and gaming apps are less preferred.

Smartphone usage patterns also vary markedly across time and between parks, reflecting differences in weekday versus weekend preferences, as well as park location and function. Parks with higher weekday smartphone use are typically centrally located and situated near workplaces or educational institutions. In contrast, parks with greater usage on weekends tend to be found in residential suburbs and feature more natural landscapes. These findings are consistent with prior research on temporal patterns in urban green space usage~\cite{bertram2017differences,pinto2024temporal}.

\rev{Complementary research shows that user-generated content (UGC) can reveal important perceptual and experiential dimensions of park use. For instance, large-scale analyses of online reviews using NLP have linked positive and negative park perceptions to specific environmental features, \eg water, colorfulness, and tranquility, highlighting how design attributes shape reported experiences across cultural contexts \cite{huai2022environmental}. More broadly, user-generated geographic information has been shown to provide insight into where, when, and how people use and value urban green spaces, while also emphasizing platform-specific biases and the value of combining multiple digital traces \cite{Heikinheimo2020Understanding}. Social media user attributes further demonstrate how differences between user groups can be inferred from what people share and where they visit, offering a window into heterogeneous human–environment interactions in parks \cite{Song2020Using}. While these UGC-based approaches capture stated perceptions and meanings expressed through text and images, our passively collected app-level mobile traffic provides a complementary lens on revealed behavior at scale, passively capturing continuous temporal rhythms and functional differentiation across parks, including everyday routines that may be underrepresented in voluntary posting and review activity.}

% ---
% \subsubsection*{Functional Typologies and Socio-Spatial Dynamics of Urban Park Usage}
\subsubsection*{Mobile Traffic Reveals Two Spatial Theories for Urban Parks}
% ---

% Mobile traffic consumption and application preferences enable the identification of three primary functional uses of parks: cultural, lunchbreak (work-related), and recreational\rev{, each strongly associated with the geographic and social contexts of their surroundings.}
% %
% This variability of smartphone usage allows us to test two hypotheses---sometimes conflicting, sometimes complementary---about how urban parks are shaped by economic pressure and social practices.

\rev{Mobile traffic consumption and application preferences provide empirical support for both hypotheses (\autoref{sec:clusters_socioeconom}), showing how economic pressure and socio-spatial context jointly shape park multifunctionality.}

\rev{Our findings support the \textit{Central-City hypothesis}, in the sense that social media apps are more commonly used in both \textit{Cultural and Lunchbreak parks}, which are centrally located, historically significant, and attract the highest smartphone activity. These parks show marked temporal shifts in app usage: social media and productivity apps are favored on weekdays but decline on weekends. In contrast, Recreational parks show steadier patterns: app preferences on weekends in both central park clusters move closer to those seen in Recreational parks (details about application preference variations between weekdays and weekends in Supplementary Information), reflecting increased use of fitness and recreational apps. This flexibility highlights the adaptive, multipurpose role of central parks, in areas with workplaces, tourist attractions, and residential functions.
Meanwhile, toward the \textit{Socio-Spatial hypothesis}, suburban parks are typically larger and more nature-oriented, aligning with our observations of \textit{Recreational parks}, which see lower overall traffic but higher weekend use. These parks display distinctive digital behavior aligned with everyday leisure and physical activity, consistent with research indicating that park visitors engage in more physical activity on weekends~\cite{fontan2021active}.}

Building on these hypotheses, we find that suburban parks exhibit \textit{temporally distributed multi-functionality} shaped by neighborhood socioeconomic context and digital cultures. Parks located in lower-income neighborhoods show greater preference for passive digital recreation such as gaming, while higher-income  neighborhoods  display greater use of fitness tracking and productivity apps, consistent with the lifestyle segmentation theory~\cite{florida2019rise}.
% , which posits that values such as health, flexibility, and visibility shape public space usage. In this perspective, park use reflects not just urban morphology but also the lived experiences of the surrounding community.
%
\rev{In contrast, centrally located Cultural and Lunchbreak parks, situated in wealthier and more socioeconomically diverse areas, display the highest app diversity and strongest weekday-weekend shifts, reinforcing the coexistence of top-down spatial pressures and bottom-up social practices in shaping park usage given the context in which they are embedded.}

% This pattern aligns with the \textit{Social-spatial hypothesis}, as suburban parks—generally farther from the city center and in areas marked by lower median income and more homogeneous populations—exhibit more uniform, “single-use” mobile application behaviors. Their recreational emphasis and weekend peaks reflect a bottom-up, collective appropriation driven by the area’s social and demographic fabric.
% %
% Conversely, Cultural and Lunchbreak parks are primarily located near the urban core, in neighborhoods characterized by higher income levels and greater income inequality. The socioeconomic diversity of these central areas likely fuels the richer and more multifaceted patterns of smartphone usage observed. This supports the \textit{Central-City hypothesis}: parks in dense, expensive central neighborhoods must serve diverse populations and function in multiple ways to respond to spatial and economic pressures.

\rev{Our findings echo previous empirical studies. In Berlin, users favored nearby parks on weekdays irrespective of their size but traveled farther on weekends to large parks with group amenities~\cite{bertram2017differences}.
% This is echoed in our observation that centrally located Lunchbreak parks experience heightened weekday usage, while larger suburban recreational parks see more weekend visitors. 
Similarly, in Coimbra accessibility and demographics shape park activities, with central parks attracting younger and more socially oriented uses~\cite{pinto2021environmental}. 
Our analysis on mobile data reproduces these patterns, showing intensified weekday activities in central parks and weekend-oriented recreational use in suburban ones.}
% These sociodemographic and behavioral contrasts are reflected in our mobile data, as central parks demonstrate higher smartphone usage and greater app diversity compared to the primarily weekend-recreational activity observed in suburban parks.

% ---
\subsubsection*{Smartphone Utilization Guiding Better Urban Park Planning}

\rev{The insights from this research can substantially inform urban planning practice and advance our understanding of citizen behavior in parks and other public spaces \cite{Li2023Multi}. Large-scale urban sensing is increasingly recognized as a key component of data-driven and evidence-based planning in contemporary cities \cite{batty2018digital, cesario2023big}. By leveraging mobile network data, planners and policymakers gain scalable, fine-grained tools to passively monitor park usage over time and identify behavioral patterns that traditional surveys or static datasets may miss \cite{Yuan2024UniST, mears2021mapping}, leveraging already existing infrastructures from mobile operators. This capability supports more responsive and adaptive park management, guiding targeted interventions in infrastructure, programming, and resource allocation \cite{Groenewegen2006Vitamin, dietz2024examininginequalityparkquality}.}

\rev{Unlike conventional visitation metrics that capture only presence, app-level mobile traffic provides insight into functional patterns of park use. Detecting under-use or single-functionality through smartphone data can reveal cases where parks may require investment in new amenities or programs tailored to diverse socioeconomic and cultural groups \cite{Hui2023Large, park2025associations}. Such evidence-based approaches help promote equity and inclusion, both key goals of sustainable urban development and environmental justice \cite{suarez2024exploring}. They also allow clearer evaluation of whether these investments reach and support the intended communities over time, and help avoid unintended effects such as gentrification and greater socio-spatial inequality \cite{wolch2014urban}.
Furthermore, the methodological framework established here can extend beyond parks to other urban green spaces and domains, supporting data-driven planning for biodiversity, climate resilience, and well-being across cities \cite{Hui2023Large, kifayatullah2025equitable}.}

\rev{Comparable methodological debates have emerged in transportation research, where passively collected smartphone and mobile network data have been proposed as complements to traditional travel surveys \cite{nitsche2014supporting, harding2021we}. Rather than replacing established survey instruments, passive data enhance them by enabling continuous measurement and reducing respondent burden. Harding et al. \cite{harding2021we} note that app-level tracking (\eg GPS pings) may impose battery and compliance burdens that affect data quality. Mobile network probes can be a less invasive alternative to device-level applications and have been shown to approximate origin–destination matrices derived from household surveys \cite{bonnel2018origin}, though such approaches require careful validation of representativeness given operator coverage. Our work builds on this logic by leveraging aggregate mobile network traffic as a less intrusive alternative to app-level tracking while acknowledging similar representativeness considerations.}

\rev{In parallel, smartphone-based ecological momentary assessment  studies demonstrate the potential of mobile surveys to reach larger and more diverse samples than traditional in-person instruments \cite{de2021smartphone}. However, active self-report approaches remain vulnerable to sampling and participation biases, including overrepresentation of specific demographic groups \cite{saleh2017examining}. Moreover, developing and maintaining bespoke survey applications requires significant technical and financial resources \cite{mcewan2020shmapped}. Participatory tools—including surveys, chatbots, and civic engagement platforms—provide rich insight into users’ motivations, perceptions, and satisfaction, but they are typically episodic and rely on self-selected respondents.}

\rev{Passive app-based inference differs fundamentally in that it captures revealed behavior at scale, offering continuous population-level signals of when and how parks are used, identifying patterns of activity, temporal rhythms, and functional differentiation across parks. Passive sensing can provide a behavioral baseline to participatory engagement surveys, helping interpret self-reported experiences.
A hybrid planning model therefore offers the most promising path forward. Large-scale app-based inference can first identify parks exhibiting unusual, inequitable, or rapidly shifting usage patterns. Planners can then deploy targeted surveys, chatbot-based engagement, or participatory workshops to investigate underlying motivations and user needs in those specific contexts. By combining continuous behavioral monitoring with context-sensitive engagement tools, cities can move toward more adaptive, evidence-based, and equitable park management.
Importantly, when leveraging large-scale passive measurements, researchers must be careful with properly assessing the representation of the population of the mobile operator versus the population reported in Census surveys, with careful consideration to avoid under-representation of certain shares that may not actively engage with smartphones.} 

\rev{Finally, we find that our data-driven park types partially align with existing park taxonomies.
There is a large agreement in broad functional categories, i.e., cultural, community/lunchbreak, and recreational types, while several details cannot be captured using mobile data alone, leading to missing distinctions such as vegetation-based subtypes\cite{ibes2015multi} or management-driven categories that depend on physical features and institutional structures rather than visitation patterns.
In Beijing, parks are classified by the Beijing Gardening and Greening Bureau into five types (``comprehensive'', ``cultural relic'', ``ecological'', ``recreational'', and ``community'') based on dominant functions and management structures \cite{ZHANG201827,KONG2022104482}. Our Cultural parks correspond closely to cultural relic parks, as both integrate historical landmarks within green spaces. The Lunchbreak category resembles community parks that primarily serve nearby users for daily recreational activities. Meanwhile, Recreational parks overlap with comprehensive, ecological, and recreational types due to their stronger emphasis on nature-oriented features.
Similarly, a feature-based classification in Phoenix identified five park types: ``Suburban Amenity,'' ``Green Mini,'' ``Native Desert Preserves,'' ``Green Neighborhood,'' and ``Urban Core''\cite{ibes2015multi}. While we see overlaps with our Lunchbreak and Recreational types, they do not clearly capture our Cultural class. This difference likely stems from their reliance on vegetation and general amenities as clustering features, without distinguishing historical landmarks. Indeed, a distinct Cultural category may not emerge in all cities, as the presence of iconic heritage sites varies across contexts.
In Tokyo, parks were classified into six temporal types based on mobile phone data \cite{ren2026planning}: ``Event-oriented,'' ``Weekend-centric,'' ``Commuter-accessible,'' ``Daily-leisure,'' ``Balanced-usage,'' and ``Low-intensity.'' Their temporally driven approach parallels ours, particularly through the use of weekday–weekend dynamics, while the integration of holiday effects in Tokyo adds a different level of discernment.
Smartphone datasets linked to mobility can also aid in clustering user-level patterns, as was done for peri-urban parks in Tokyo, where mobile indicators revealed 4 distinct groups: ``Park explorers,'' ``Regulars,'' ``Park enthusiasts,'' and ``Staff''~\cite{guan2024exploring}. Future integrations between datasets that allow for user- and park-level classification can help further strengthen the link between the utilization of parks based on its physical features~\cite{dietz2024examininginequalityparkquality} and studies that classify them based on users engagement across different granularity. 
}

% ---
\subsubsection*{\newrev{Behavioral Signatures as Proxies for Park Multifunctionality: Urban Planning and Equity Implications}}
% ---

\newrev{Mobile app traffic patterns provide a novel, scalable proxy for understanding park multifunctionality and identifying equity disparities in park use. While our data cannot directly measure health outcomes, prior literature documents that park multifunctionality and park-based interventions promote physical and mental health, social cohesion, and well-being~\cite{dietz2024examininginequalityparkquality}. The accurate characterization of park functions from app traffic offers potential to enhance park design and resource allocation. Mobile application signature analysis may inform targeted improvements to park amenities and programming, and showcases its potential as a data source for future work jointly exploring how screen time and green time relate to urban health\cite{oswald2020psychological}, with the goal of reducing screen time.}
\newrev{This approach is operationally valuable: policymakers and urban planners can use these functional signatures, passively collected by mobile network operators, to inform infrastructure and programming decisions, especially when combined with information on provision of different amenities within the parks. For example, Recreational parks showing low physical activity app engagement could be candidates for fitness facilities or sports programming, while parks with high gaming and social media use could receive community and cultural programming. Future research combining app-level characterization with health surveys could empirically validate whether these functional signatures correlate with individual health outcomes.}

% Beyond urban planning applications, smartphone utilization patterns reveal how urban parks can impact public health. The accurate characterization of park multifunctionality enables evidence-based interventions to maximize the health benefits of these spaces.
%
% Park-based interventions promote physical and mental health, social cohesion, and well-being~\cite{dietz2024examininginequalityparkquality}, making them important spaces for health promotion.
%
% Mobile application traffic signatures enable identification of parks with distinctive usage patterns. They reveal not just where users are present, but what activities they engage in and how park design shapes these behaviors. 
%
% This information is actionable: policymakers and public health practitioners can use these functional signatures to target interventions, especially when combined with information on provision of different amenities within the parks~\cite{dietz2024examininginequalityparkquality}. For example, recreational parks showing low physical activity app engagement could receive fitness facilities or sports programming, while parks with high gaming and social media use could receive community and cultural programming through social prescribing initiatives, thus, supporting health-related activities across diverse park types in metropolitan areas.

\newrev{The socio-spatial findings linking suburban park app signatures to neighborhood income reveal disparities in park use across income levels in the built environment. Substantial evidence demonstrates that green space availability inversely correlates with socioeconomic deprivation~\cite{ngan2025inequality}, suggesting equity-motivated investment in parks serving lower-income areas. Our finding that suburban Recreational parks reflect neighborhood socioeconomic characteristics suggests that app-based characterization could identify under-served communities for targeted park improvement or programmatic intervention. However, whether such interventions achieve health improvements remains an open question requiring longitudinal or intervention studies.}

% The Socio-Spatial findings linking suburban park app signatures to neighborhood income have direct implications for health equity in the built environment. Substantial evidence demonstrates that green space availability inversely correlates with socioeconomic deprivation, and that green space interventions show greater health impact in deprived neighborhoods than affluent areas \cite{ngan2025inequality}. Prioritizing park-based health interventions in lower-income areas thus aligns with both equity principles and public health effectiveness. Our finding that suburban recreational parks reflect neighborhood socioeconomic characteristics suggests that app-based characterization can identify communities where park-based interventions would have maximal health impact. Additionally, this methodology addresses a measurement gap: passively collected digital data enables real-time, large-scale monitoring of park utilization patterns without the resource constraints and reactivity limitations of traditional surveys, facilitating rapid identification of disparities and evaluation of intervention effectiveness \cite{jay2022use}.

\newrev{Mobile application traffic data may offer a novel approach to monitoring temporal shifts in park visitation, which could inform climate-responsive park design. Prior research demonstrates that extreme temperatures reduce park visitation~\cite{kabisch2020physical,qin2021thermal}, and that this effect disproportionately affects vulnerable groups~\cite{wang2026city}. The integration of weather data with app consumption could enable future systems for monitoring park usage and climate conditions, allowing planners to identify parks where heat mitigation interventions (shade provision, tree canopy expansion, cooling infrastructure) may be most needed through the joint analysis of variation in temperature and specific key application categories, e.g., a potential reduction in Fitness application usage and an increase in Social Media usage in parks connected to a rise in temperature.}

\newrev{Our methodology demonstrates that passively collected digital data can characterize the functions of medium to large sized parks at scale, enabling policymakers to identify disparities in park use across communities. This approach offers a cost-effective, scalable tool for informing evidence-based allocation of park resources and programming to under-served neighborhoods. Future studies combining this approach with health surveys and intervention trials could establish whether such digitally-informed improvements translate to measurable health equity gains.}

% Our methodology demonstrates that passively collected digital data can characterize park functions at scale, enabling policymakers to identify priority communities for targeted interventions. This approach offers a cost-effective, sustainable tool for evidence-based modification of the built environment to promote health equity in under-served neighborhoods.

%------------------------
%------------------------
\subsection{Limitations and Future Directions}
%------------------------
%------------------------

\newrev{Our study is not without limitations. First, our quality metrics exclude small and pocket parks because the spatial resolution of mobile traffic data does not reliably separate activity inside these parks from surrounding areas. In addition, uneven base-station density can introduce assignment error, particularly in less densely deployed areas, and our coverage filter mitigates this at the cost of excluding some locations.}
% Our study is not without limitations. First, the quality metrics excluded small and pocket parks due to  constraints of the data source: the spatial resolution of mobile traffic data does not allow us to fully isolate activity within small-scale parks from that in surrounding areas. 
%
% \newrev{Furthermore, the distribution and density of base stations can introduce estimation errors in visitor traffic assignment, especially in less dense areas. While our quality of coverage methodology helps mitigate this issue by filtering parks with an insufficient overlap with base stations (mainly small and pocket parks), we acknowledge that this limits the analysis of areas with a less dense deployment.}
%
As mobile networks continue to densify, in the future, it may become feasible to include more and more smaller parks using the same methods, \newrev{as well as further explore areas that currently have a less dense deployment}.

Second, our approach relies on mobile network deployments for geolocation, \rev{with traffic aggregated at antenna level, preventing inference of visitor origins (\eg home neighborhoods) and separation of local residents from inbound visitors. Future work combining higher-resolution mobility traces (\eg GPS) with app-level activity could directly quantify mobility flows and stratify behaviors by visitor origin.}
However, accessing high-resolution mobility data typically requires linking information across institutions and potentially identifying individuals, which poses significant privacy concerns and limits scalability, as it conflicts with current data protection regulations.

It is also important to note the limitations associated with the collection of app-level traffic data. Although measurements are passively conducted via the network operator, they still require each application to have a distinct \textit{traffic footprint}. For example, our Gaming category includes titles such as Pokémon Go and Fortnite, which have a significant online component and may be biased toward certain demographics. In contrast, games that generate little to no network traffic may not be well represented by this method (\eg Candy Crush).
\newrev{Additionally, the 35\% operator market share may not fully represent digital habits across all demographic groups, including some elderly and lower-income populations, which may under-represent certain usage patterns. Future work could mitigate this through demographic weighting or by integrating complementary data sources.}
% \newrev{Additionally, the 35\% market share of mobile traffic used in this study may not fully represent the digital habits of elderly or lower-income populations, who may not use data-intensive apps as frequently. This could lead to under-representation of certain demographic groups in our analysis, particularly in Recreational parks that serve these populations. Future research could explore methods to adjust for these potential biases, such as weighting traffic data based on demographic distributions or integrating additional data sources.}

Lastly, our analysis is limited to parks within the metropolitan area of Paris, potentially biasing the findings toward the spatial, cultural, and socioeconomic characteristics of this region. Future research could extend this framework to other French cities to assess national generalizability and, more broadly, to diverse global urban contexts to explore how relationships between smartphone usage and park activity may vary across sociocultural and infrastructural settings.

%% file: sections/4-methods.tex
%--
\subsection{Mobile Traffic Data Collection}
\label{subsec:data_collection}
%--

This study builds on real-world measurements from the network of a major mobile operator in France, which holds a 35\% market share and serves over 20 million customers in the country. 
The data collection period spans from January 31 to May 31, 2023, capturing traffic from 4G and 5G non-standalone radio access technologies. This traffic was collected through passive measurement probes at approximately 25,000 antennas in the Paris metropolitan area.
To confirm that the traffic is stable over the study period, we performed an Augmented Dickey--Fuller test on each park, and found that it was stationary; that is, there was no significant change in volumes over the studied period.
%The data is available on aggregated basis in 1-hours... calculating statistics about all traffic sessions of the users in both uplink and downlink directions. 
%The characterization and overall statistics of those sessions are unique to the way each mobile application is programmed by its developers~\cite{zanella2023characterizing}, and also tied to the spatiotemporal context~\cite{marquez2017not}.

User traffic is aggregated at the antenna level into one-hour intervals to protect user privacy and ensure compliance with applicable data protection laws~\cite{gdpr}. Antennas serving fewer than 10 unique users on a given day are excluded from the analysis, resulting in more than 492 million records. The data format and volumes are detailed in \autoref{tab:data_structure}.

% \fromA{Moved this to the ethics section on the main file}
% \noindent\textbf{Ethical and Privacy Considerations:} 
% %
% The data obtained from this collection was collected by the operator for network management and research purposes, and temporarily stored and processed within a secure platform at their own premises, in full compliance with Article 89 of the General Data Protection Regulation (GDPR)~\cite{gdpr}. 
% The data collection and processing were approved by the Data Protection Officer of the operator.
% The researchers involved in this study only had access to such aggregates, whose spatiotemporal resolution ensures that no data subject can be re-identified from the data, which does not constitute personal data in terms of GDPR.

%The different mobile applications, identified using Deep Packet Inspection (DPI), follow an imbalanced traffic distribution following the Pareto principle, where approximately $80\%$ of traffic is generated by the top $20\%$ applications~\cite{zanella2023characterizing}.
%To obtain meaningful insights into traffic consumption in parks, we started with the 100 most popular apps and excluded any non-insightful apps like App Store traffic and generic network protocols, which are not linked to any specific activity. The final selection of 41 apps account for 65.6\% of total traffic (79.4\% excluding network protocols), representing the majority of application usage in the studied area.

%--
\subsection{Selecting Mobile Applications}
\label{subsec:mobile_apps}
%--

The mobile applications are identified through a proprietary Deep Packet Inspection tool owned by the operator. It analyzes packet headers to determine the application they belong to, but does not access packet payload content. 
Applications exhibit an imbalanced traffic distribution that follows the Pareto principle with approximately 80\% of traffic being generated by the top 20\% of applications, an expected traffic pattern~\cite{zanella2023characterizing}.
To derive meaningful insights into traffic consumption in parks, we started with the 100 most popular apps and excluded non-informative categories such as App Store traffic, unidentified encrypted traffic, and generic network protocols, which are not directly linked to specific user activities. The final selection of 41 applications accounts for 65.6\% of the total traffic the mobile network sees (79.4\% when excluding encrypted traffic and network protocols), capturing the majority of application usage in the studied area.
An expert panel was used to categorize individual apps into high-level categories. Their task was to group applications together by the activity the user is engaged in when using these apps. Any conflicts were resolved in a subsequent discussion phase, which also included naming the resulting categories: fitness, games, music, news \& information, productivity, shopping, social, travel, and video. The applications included in each category can be seen in \autoref{tab:apps}.

\begin{table}[t]
    \centering
    \caption{Overview of the dataset and the structure of the collected traffic data, aggregated with 1 hour granularity. Overall 492 million records were analyzed.}
    \small
    \begin{tabular}{rrll}
    \toprule
    &\textbf{Study Area} & Metropolis of Paris & 7 million citizens \\
    &\textbf{Study Period} & Jan 31st - May 31st 2023 & 4 months \\
   & \textbf{Data Source} & \multicolumn{2}{l}{A major French mobile network operator} \\
    \midrule
    \multirow{5}{*}{\rotatebox{90}{\textbf{Structure}}} &
%\multicolumn{3}{c}{\textbf{Record Structure}}\\
        \textbf{Antenna ID} & 16-bit unique antenna ID & 25,000 antennas \\
        &\textbf{Application ID} & 5-digit app ID & 41 apps\\
       & \textbf{Timestamp} & Unix timestamp & hour\\
       & \textbf{Downlink} & Aggregated traffic volume & bytes\\
        &\textbf{Uplink} &Aggregated traffic volume & bytes \\
        \bottomrule
    \end{tabular}
    \label{tab:data_structure}
\end{table}

\begin{table*}[h]
    \centering
    \caption{Studied mobile applications for their usage within urban parks and their respective categories.}
    \label{tab:apps}
    \begin{tabular}{p{1.552cm}l}
    \toprule
        \textbf{Category} & \textbf{Mobile Application} \\ 
        \midrule 
        Fitness & Garmin Connect \\ 
        Games & Pokémon Go, Fortnite \\
        Music & Apple Music, Spotify, Deezer, Soundcloud \\
        News & Wikipedia, Sports News, NewsPaper, NewsMag, Weather, Tripadvisor \\
        Productivity & Skype, Microsoft Mail, Google Drive, Gmail, Finances, Dropbox \\
        Shopping & Amazon \\ 
        Social & Instagram, WhatsApp, Facebook, SnapChat, LinkedIn, Pinterest, Twitter, Facebook Messenger, TikTok \\
        Travel & Uber, Waze, Airfrance, Transport \\
        Video & Twitch, Periscope, Youtube, Netflix, Molotov TV, DailyMotion, Apple Video, Facebook Live \\
        \bottomrule
    \end{tabular}
\end{table*}

% --
\subsection{Quantifying the Quality of Coverage in Parks}
\label{subsec:park_selection_methodology}
% --

%%%%%%%%%%%%%%%%%%%%%%%%%%%%%%%%
% Quality of coverage metrics
%%%%%%%%%%%%%%%%%%%%%%%%%%%%%%%%

Our methodology for selecting parks based on the quality of the antennas' coverage is next detailed. This is done in three steps, each of them comparing the geometric overlap of Voronoi and park polygons. In the following, we use `illumination' to refer to the area of the park that is serviced by a specific antenna. These steps are partially visualized in \autoref{fig:park_selection_methodology}A-C:

\noindent\textbf{Step 1: Antenna selection.} 
Considering $A_v$ as the expected Voronoi coverage area of antenna $v$, and $A_p$ as the area of park $p$, we defined the intersecting illumination area as 
$A_{i,v} = A_v \cap A_p$.
Consequently, the illuminating ratio of an antenna was defined as  
$I_{pv} = \frac{A_{i,v}}{A_v}.$  
In the first step, we selected all antennas which have an $I_{pv} \geq \alpha$ with any park $p$, where $\alpha = 0.1$, excluding those below this threshold.
This step was mostly required to select antennas that overlap with parks, thus, a small $\alpha$ was sufficient for our purposes.
For each park $p$, the selected set of Voronoi cells is denoted as $V_{sel,p}$. Aggregating this set from the perspective of a park $p$, we obtained the coverage precision as: 

\begin{equation}
    CP_{p} = \sum_{\substack{v \in V_{sel,p}}} \frac{A_{i,v}}{A_{v}}
\end{equation}

\noindent As a result, $1,141$ parks were identified as not being serviced by any antenna with sufficient illumination ($V_{sel,p} = \emptyset$). These parks, shown in gray in \autoref{fig:park_selection_methodology}C, were predominantly small green spaces such as roadside strips or pocket parks, e.g., the park shown in \autoref{fig:park_selection_methodology}B2. These are significantly smaller than the Voronoi cells, making traffic assignment uncertain.  
%In contrast, larger parks, such as the one of \autoref{fig:park_selection_methodology} (top-left), contain more Voronoi cells with higher coverage ratios (in green).
%Antennas with minimal intersection (in red) are excluded.  

\noindent\textbf{Step 2: Park area coverage.} 
For the parks where $V_{sel,p} \neq \emptyset$, we verified the percentage of $A_{p}$ covered by the selected antennas.  
We defined this as the park illumination, calculated as:  

\begin{equation}
I_{p} = \frac{\sum_{\substack{v \in V_{sel}}} A_{i,v}}{A_{p}}
\end{equation}

\noindent We retained all parks where $I_{p} > \beta$, with $\beta = 0.8$. This threshold ensures that the majority of each park's area is covered, preventing the inclusion of parks with only a small portion adequately covered by antennas in the analysis.
A total of $76$ additional parks were excluded at this stage, marked in red in \autoref{fig:park_selection_methodology}C.
In the example park in \autoref{fig:park_selection_methodology}B1, one can see almost all of the park area is illuminated by selected Voronoi cells (in green).

\noindent\textbf{Step 3: Coverage quality.} 
Finally, we applied a coverage quality metric $Q_{p}$ for each park $p$, which balances the ratio of selected Voronoi cells inside the park with how much area is covered by antennas deemed with good coverage. 
$Q_{p}$ is inspired by the F-score and is defined as: 

\begin{equation}
    % Q_{p} = 2 \times \frac{CP{p} \times I{p}}{CP{p}+I{p}}
    Q_{p} = \frac{2 \cdot CP_{p} \cdot I_{p}}{CP_{p} + I_{p}}.
\end{equation}

\noindent Applying a threshold of $Q_{p} \geq \gamma$, where $\gamma =0.4$, an additional $78$ parks were excluded, marked yellow in \autoref{fig:park_selection_methodology}C.
To determine $\gamma$, we observed the trade-off between the statistics of the different coverage metrics and the number of parks that can be analyzed. 
%\autoref{fig:park_selection_methodology} illustrates the final outcome of applying these steps to parks in Paris.  
While the choice of thresholds for $[\alpha, \beta,\gamma]$ can be adjusted to satisfy the needs of different use cases, we are confident that the 45 selected parks (detailed in the Supplementary Information) have reliable coverage from mobile traffic antennas while minimizing the inclusion of traffic from surrounding neighborhoods.

% --
\subsection{Accuracy Analysis of Park Selection}
\label{subsec:acc_analysis}
% --
Obtaining ground truth for base station coverage is challenging and inherently complex to model~\cite{deepmend_secon}, largely because Voronoi approximations already deviate from the physical realities of wireless signaling.
Our method strikes a balance between the inevitable error introduced by Voronoi approximations and the number of parks that can be analyzed.
This approach enables an informed selection of spaces that can be reliably studied, while being able to quantify the error within the framework of Voronoi approximation.
\autoref{tab:traffic_assignment} compares traffic assignment methods. Applying our antenna-based method on the 45 selected parks, we get a median $Q_p$ of $0.529$, a $33.26\%$ increase compared to the cell-based method.
Conversely, if we applied the three steps with the same thresholds but using cell tower Voronoi cells, it would only be possible to analyze 24 instead of 45 parks.

%\adam{Is there anyway we could discuss the validity of the results? What's the ``baseline'' for doing this quantification?}
%\linus{fill and finalize discussion of \autoref{tab:traffic_assignment}}
%\andre{The test between traditional and shifted voronois did not result in much change in the metrics themselves ($4\%$ improvement). However, the number of selected parks went from 21 to 45. This should be the justification.}

\begin{table}[t]
    \centering
        \caption{Accuracy evaluation of park traffic assignment. Our antenna-based method is 33.26\% more accurate compared to the standard cell tower-based traffic assignment. The tabulated values are the medians.
        % \andre{As we cannot test the traffic of the grid to showcase the lost of details, I recommend dropping from there and citing the paper on related works that mention this. Otherwise it's a bad look for us.}
        }
        \label{tab:traffic_assignment}
    \begin{tabular}{rlll}
    \toprule
         Method  & $CP_p$ & $I_p$ & $Q_p$\\
    \midrule
         % Grid-based (large) & 0.350 & 0.951 & 0.507 & 59 \\
         % Grid-based (medium) & 0.383 & 0.965 & 0.544 & 179 \\
         % Grid-based (small) & 0.386 & 0.956 & 0.550 & 700 \\
         Base station-based & 0.262 & 0.894 & 0.399  \\
         Antenna-based & 0.367 & 0.947 & 0.529 \\
    \bottomrule
    \end{tabular}
\end{table}

% % --
% \subsection{Converting antenna-oriented traffic to park-oriented}
% % --
% \linus{Not sure, whether it would just be sufficient to mention this in one sentence at the end of the previous section.}

% --
\subsection{Converting Antenna-oriented Traffic to Park-oriented}
\label{subsec:traffic_conversion}
% --
With the selected set of parks and their respective antennas with good coverage, the final step of data processing involves converting the traffic $T_{v}(t)$, originally associated with Voronoi geometries, into the traffic $T_{p}(t)$ observed within each park $p$.  
This is achieved by aggregating the traffic from all selected antennas in $V_{sel,p}$, weighted by their respective illumination ratios:  

\begin{equation}
    T_{p}(t) = \sum_{v \in V_{sel,p}} I_{pv} \cdot T_{v}(t)
\end{equation}

Thus, the traffic of park $p$ at each time instant $t$ is given by the weighted sum of the traffic from all its selected Voronoi cells in $V_{sel,p}$, where the weights correspond to their respective illumination ratios $I_{pv}$.  

% --
\subsection{Measuring Mobile App Preferences}
\label{subsec:rsca}
% --

To quantify which app categories are used more in parks versus other parts of the city, we need a metric that captures relative traffic consumption across different areas. The simplest method would compare absolute traffic volumes, but this approach is problematic because different app types have very different traffic demands (e.g., video streaming uses much more data than instant messaging). Instead, we use the Revealed Comparative Advantage~(RCA)~\cite{RCA} metric, which has been successfully applied in mobile traffic analysis and clustering~\cite{mishra2022second,bakirtzis2023characterizing,zanella2024characterizing}. It defines the level of over- or under-utilization of a specific mobile application at a given antenna relative to the entire set of applications and antennas.

For a mobile application \(a \in A\), and an antenna \(v \in V\), the RCA is defined as:
\begin{equation}
     RCA_{a,v} = \frac{T_{a,v}/\sum_{a' \in A}T_{a'v}}{\sum_{v' \in V}T_{av'}/\sum_{a' \in A,v' \in V}T_{a'v'}}, 
    \label{eq:RCA}
\end{equation}
\noindent where \(T_{a,v}\) is the traffic for application \(a\) at antenna \(v\); $\sum_{a' \in A}T_{a'v}$ is the total traffic from all applications at antenna \(v\); $\sum_{v' \in V}T_{av'}$ is the total traffic for application \(a\) across all antennas; and $\sum_{a' \in A,v' \in V}T_{a'v'}$ is the overall traffic for all applications and antennas in the network.
Values of \(RCA\) above or below 1 indicate that an antenna has more or less consumption of a certain application compared to other antennas in the network. A downside of \(RCA\) is that while its lower boundary is 0, the upper boundary is \(\infty\). This can unbalance the distribution of \(RCA\) values and cause issues in clustering algorithms~\cite{Symmetric_RCA}. %, Symmetric_RCA}.

To address this, we use the \ac{RSCA}~\cite{Symmetric_RCA} defined as
$$
RSCA_{a,v} = \frac{RCA_{a,v} - 1}{RCA_{a,v} + 1},
$$
\noindent which yields values bounded in the interval \(\left[-1, 1\right]\). In this scale, values above 0 indicate over-utilization, and values below 0 indicate under-utilization. This results in a better-balanced distribution of values, eliminating the long tail observed with the original \(RCA\) metric.
\rev{We calculate RSCA values over the median traffic in parks, for each application category.}

%--
\subsection{Clustering Mobile Traffic Toward Park Categories}
\label{sec:clustering}
%--

\rev{For characterizing parks, we use the \textit{ratio of traffic on weekdays versus weekends} and individual \textit{application preferences}, measured with the RSCA. These features allow us to group parks by their temporal use patterns (\eg weekday-oriented, weekend-oriented, or consistently busy) and by the types of mobile applications users favor in these spaces (\eg parks with higher use of gaming applications or lower use of social media).
Transforming raw traffic volumes into these derived features enables fair comparison across parks of different sizes. 
The weekday-to-weekend ratio captures relative temporal preference, while the RSCA shows whether an application is over- or underrepresented in a park compared with other city areas, allowing park types to be partially defined by their temporal rhythms and not just visitation volume, as, for example, larger parks or parks with more visitors naturally produce higher data volumes. By relying on relative measures, we reduce volume bias linked to park size or popularity and isolate temporal and application usage preferences.}

\rev{Concretely, we cluster all parks described by 10 features: the weekday-to-weekend ratio and the RSCA values for nine app categories (Fitness, Games, Music, News, Productivity, Shopping, Social, Travel, and Video). We apply Spectral clustering, a graph-based method that groups data points using similarity structure rather than the original feature space\cite{ng2001spectral}. The method builds a similarity graph in which each node represents a park and edges encode pairwise similarity. From this graph, we compute a Laplacian matrix and derive a low-dimensional embedding, which is then clustered to obtain the final groups. Compared to simple clustering approaches like k-means, spectral clustering works well for complex and non-convex data structures and given the number of parks in any city, the computational overhead is not relevant~\cite{von2007tutorial}. We use nearest neighbors as the affinity matrix with $n=10$ neighbors and discretization to obtain the final clusters.
We select the number of clusters using three validation metrics: the Silhouette score, the Calinski–Harabasz index, and the Davies–Bouldin  index (see Supplementary Information). We choose $k=3$, which achieves strong Silhouette and Calinski–Harabasz values (second only to $k=2$) and the best Davies–Bouldin score, while providing enough clusters for further analysis.
To further confirm cluster stability under noise perturbation, and to address sensitivity to graph density, we compute the Adjusted Rand Index (ARI) across different settings. The results remain stable under these variations (see Supplementary Information).}

%--
\subsection{Relative Importance of Flickr Tags}
\label{sec:flickr_method}
%--

To compute the relative importance of a Flickr tag in a park (details in the Supplementary Information), we first compute the probability that an image is tagged with this tag $p_{tag}$ assuming a uniform distribution:

\[
p_{tag} = \frac{|tag|}{\sum_{tag \in tags} |tag|}
\]

In the second step, we compute the expected number of occurrences of a tag in a park $\hat{n}_{\{tag|park\}}$, by multiplying the  relative frequency with the number of tags of this park.

\[
\hat{n}_{\{tag|park\}} = p_{tag} \cdot |tag \in park|
\]

Finally, we compute the ratio between the actual number of occurrences and the expected number of occurrences:

\[
r_{\{tag|park\}} = \frac{n_{\{tag|park\}}}{\hat{n}_{\{tag|park\}}}
\]

This provides us with a relative importance of each tag in each park. A value close to 1 means that this tag is occurring at an average rate, whereas values higher than 1 indicate that a tag is occurring more frequently than expected.
To clean the results, we remove English and French stop words, as well as irrelevant tags, such as ``France'', ``Paris'', ``park'', as well as Flickr-specific tags referring to photography techniques (``black \& white'', ``square format'', ``hdr'' and synonyms thereof). Finally, as it is common that in addition to the geo-coordinates, the images get tagged with the name of a location, we remove all tags that have a value above 1 in exactly one park (e.g., ``jardinduluxembourg'', ``champdemars'', ``jardindesplantes''). Note that ``toureiffel'' (and synonyms) are not removed in this step, as they have a relative importance over 1 in both Champ de Mars and in the Jardins du Trocadéro.

%% file: sections/5-appendix.tex
% --
\section{Studied Parks}
\label{subsec:park_details}
% --

We present in \autoref{tab:selected_parks} the complete set of parks included in our study, which we are able to confidently isolate and assign mobile traffic demands, following the methodology described in the methodology (Section 4.3 of the main text).
We include as well the clusters they are assigned in our analysis (Section 2.3 of the main text), their area, the number of antennas with good illumination in each, their Quality of Coverage ($Q_p$) metric and a traffic score, which is normalized across all studied parks to give a notion of traffic volume across them. 

\begin{table*}[!t]
    \centering
    \caption{The selected set of $45$ parks and their respective clusters, with their area (in ha), the number of antennas with good coverage used for their analysis, as well as the Quality of Coverage $Q_{p}$ score. We also present the normalized traffic score of each park, which adds a notion of the different volume of overall traffic at each park.
    }
    \label{tab:selected_parks}
    \begin{tabular}{m{20em} m{6em} m{5em} m{4em} m{2em} m{6em}}
    \toprule
        \textbf{Park name} & \textbf{Cluster} & \textbf{Area (ha)} & \textbf{Antennas} & \textbf{$Q_{p}$} & \textbf{Traffic score} \\
        \midrule
        Bois de Boulogne & Cultural & 836.83 & 39 & 0.85 & 0.3111 \\
        Bois de Vincennes & Cultural & 1000.28 & 42 & 0.81 & 0.3679 \\
        Champ de Mars & Cultural & 28.77 & 8 & 0.55 & 0.4804 \\
        Cité Universitaire - Parc Est & Cultural & 26.74 & 7 & 0.41 & 0.3731 \\
        Cité Universitaire - Parc Ouest & Cultural & 7.62 & 3 & 0.48 & 0.3562 \\
        Domaine national de Saint-Cloud & Cultural & 376.95 & 17 & 0.66 & 0.1614 \\
        Jardin des Tuileries & Cultural & 22.38 & 7 & 0.47 & 0.4920 \\
        Jardin du Ranelagh & Cultural & 10.14 & 5 & 0.50 & 0.3559 \\
        Parc Georges Brassens & Cultural & 7.58 & 2 & 0.55 & 0.5803 \\
        Parc Montsouris & Cultural & 16.28 & 3 & 0.61 & 0.4020 \\
        Parc Pierre-Lagravère & Cultural & 29.98 & 5 & 0.40 & 0.9471 \\
        Parc de Bercy & Cultural & 14.65 & 4 & 0.53 & 0.5322 \\
        Parc de la Villette & Cultural & 33.66 & 5 & 0.61 & 0.7282 \\
        Parc des Buttes-Chaumont & Cultural & 25.68 & 8 & 0.54 & 0.3986 \\
        Parc du Lycée Michelet & Cultural & 9.03 & 2 & 0.44 & 0.6506 \\
        Parc départemental Jean-Moulin - Les Guilands & Cultural & 25.25 & 2 & 0.53 & 0.7790 \\
        Square des Batignolles & Cultural & 1.72 & 2 & 0.47 & 0.5760 \\
        Jardin des Plantes & Lunchbreak & 16.17 & 4 & 0.42 & 1.0000 \\
        Jardin du Luxembourg & Lunchbreak & 22.23 & 7 & 0.60 & 0.3216 \\
        Jardins Abbé Pierre - Grands Moulins & Lunchbreak & 0.78 & 1 & 0.59 & 0.3486 \\
        Jardins du Trocadéro & Lunchbreak & 13.79 & 4 & 0.50 & 0.5438 \\
        Parc André Citroën & Lunchbreak & 9.58 & 3 & 0.43 & 0.4966 \\
        Parc Clichy-Batignolles Martin Luther King & Lunchbreak & 9.92 & 3 & 0.61 & 0.7398 \\
        Parc Départemental des Chanteraines & Lunchbreak & 72.06 & 4 & 0.53 & 0.2182 \\
        Parc Floral de Paris & Lunchbreak & 33.50 & 2 & 0.57 & 0.2543 \\
        Parc Monceau & Lunchbreak & 8.43 & 4 & 0.46 & 0.4667 \\
        Parc Suzanne Lenglen & Lunchbreak & 18.91 & 4 & 0.41 & 0.6661 \\
        Parc de Sceaux & Lunchbreak & 96.40 & 6 & 0.69 & 0.5154 \\
        Arboretum de Paris & Recreational & 12.98 & 1 & 0.48 & 0.0000 \\
        Parc Henri Sellier & Recreational & 25.11 & 2 & 0.46 & 0.5296 \\
        Parc Interdépartemental des Sports - Plaine Sud & Recreational & 107.80 & 5 & 0.59 & 0.0449 \\
        Parc Lefèvre & Recreational & 9.03 & 1 & 0.41 & 0.4345 \\
        Parc Nature du Plateau d'Avron & Recreational & 11.27 & 2 & 0.53 & 0.0299 \\
        Parc de la Butte du Chapeau Rouge & Recreational & 4.74 & 2 & 0.41 & 0.2519 \\
        Parc de la Vallée-aux-Loups & Recreational & 45.14 & 3 & 0.52 & 0.3782 \\
        Parc des Artistes & Recreational & 9.35 & 2 & 0.52 & 0.1407 \\
        Parc des Saules & Recreational & 6.55 & 2 & 0.42 & 0.0994 \\
        Parc départemental Georges-Valbon & Recreational & 315.63 & 13 & 0.73 & 0.1187 \\
        Parc départemental de la Fosse Maussoin & Recreational & 23.32 & 3 & 0.42 & 0.5094 \\
        Parc départemental de la Haute-Île & Recreational & 74.71 & 5 & 0.42 & 0.1999 \\
        Parc départemental de la Plage Bleue & Recreational & 40.99 & 2 & 0.42 & 0.4009 \\
        Parc départemental des Lilas & Recreational & 83.93 & 8 & 0.55 & 0.1531 \\
        Parc départemental du Sausset & Recreational & 205.30 & 9 & 0.75 & 0.1104 \\
        Parc intercommunal des Sports du Grand Godet & Recreational & 21.53 & 3 & 0.53 & 0.1094 \\
        Île de Loisirs de Créteil & Recreational & 45.86 & 7 & 0.56 & 0.3693 \\
        \bottomrule
    \end{tabular}
\end{table*}

% --
\section{Variation of RSCA on Park Clusters Across Workdays and Weekends}
\label{subsec:wdwe_rsca_clusters}
% --

We observe an interesting pattern on the RSCA values across park clusters: both Lunchbreak and Recreational parks have their application preferences change on weekends, in relation to weekdays. More specifically, categories such as News, Productivity and Social, which on Lunchbreak and Cultural parks have opposite behaviors from Recreational parks, see these patterns reverse on weekends, with preference looking closer to a Recreational cluster park. This can indicate that these parks have distinct usages across the weeks, with parks within the urban center of Paris looking more like Recreational parks during the weekend (as their work-related users may not be  present, leaving only residents of the area using those parks for more traditional recreational activities - as reflected in our measurements). 

\begin{figure*}[h]
    \centering
    \includegraphics[width=0.8\linewidth]{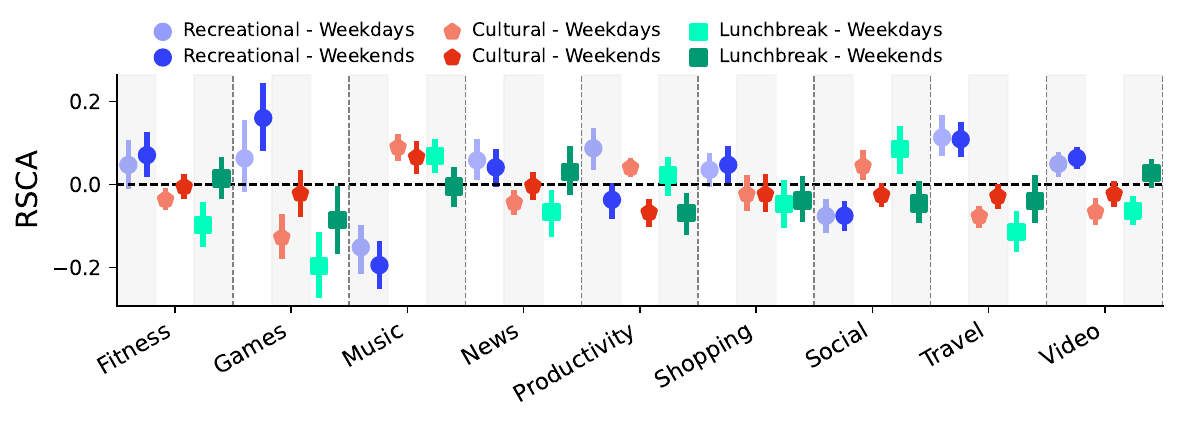}
    \caption{RSCA across apps per park category, split between weekday and weekend. Points show mean values, while lines indicate $95\%$ confidence intervals across parks in each cluster.}
    \label{fig:rsca_wdwe}
\end{figure*}

%------------------------
%------------------------
\section{Robustness and Sensitivity Analysis of Clustering}
\label{sec:cluster_robustness}
%------------------------
%------------------------

%%% Robustness
\noindent \textbf{Stability Under Noise Perturbation.}
\rev{To assess the robustness of the clustering results, we conducted a noise-perturbation stability analysis. Gaussian noise was added independently to each feature, with the noise standard deviation scaled to $1\%$, $5\%$, and $10\%$ of the empirical feature-wise standard deviation. For each noise level, spectral clustering was re-estimated with identical parameter settings. The resulting partitions were compared to the original clustering using the \ac{ARI}. 
The \ac{ARI} remained equal to $1.00$ under $1\%$ noise, indicating perfect stability, and remained high ($\text{ARI} \approx 0.94$) under both $5\%$ and $10\%$ perturbations. These results indicate that the identified park clustering is robust to moderate levels of random noise and is not driven by small fluctuations in feature values.}

%%% Sensitivity
\noindent \textbf{Sensitivity to graph density.}
\rev{To evaluate the sensitivity of the clustering results to the graph construction, we conducted a parameter robustness analysis by varying the number of nearest neighbors used to construct the affinity matrix. Specifically, we re-estimated the spectral clustering model with $n \in \{5, 10, 15, 20\}$ nearest neighbors, while keeping all other parameters fixed. The resulting partitions were compared to the baseline specification ($k=10$) with ARI values equal to $0.71$ for $n=5$, $1.00$ for $n=10$, $0.81$ for $n=15$, and $0.94$ for $n=20$. These results indicate that the clustering structure is generally stable across a reasonable range of neighborhood sizes, with higher stability observed for moderate to larger values of $n$. While smaller neighborhood sizes ($n=5$) introduce some variation in cluster assignments, the overall partition remains largely consistent, suggesting that the identified park typology is not highly sensitive to the specific choice of graph connectivity parameter.}

\begin{figure*}[h]
    \centering
    \includegraphics[width=0.95\linewidth]{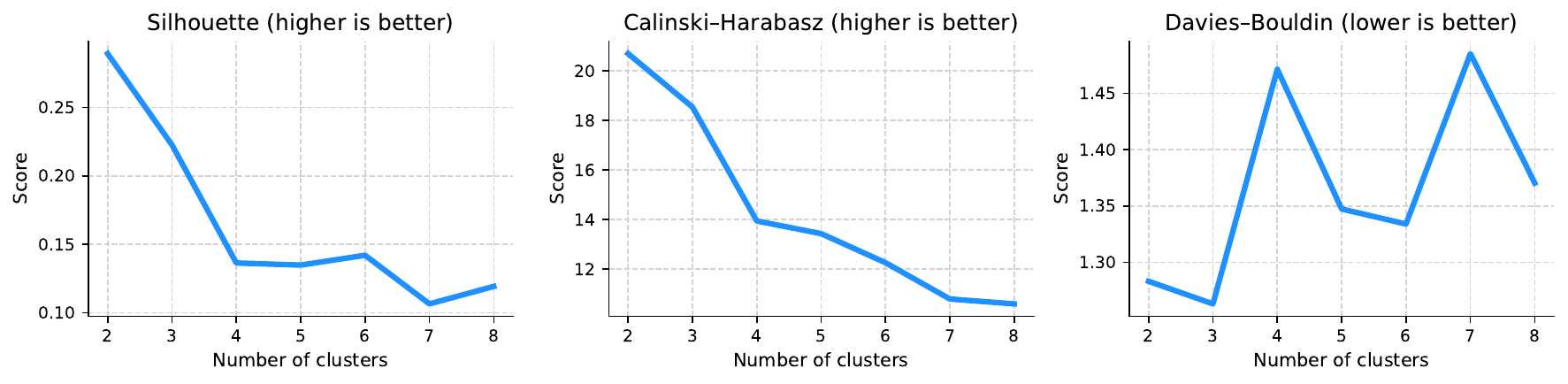}
    \caption{Comparison of internal clustering validation metrics for spectral clustering and k-means across $k=2$--$9$. Silhouette and Calinski--Harabasz scores (higher values indicate better separation) and Davies--Bouldin index (lower values indicate better separation) are reported to assess cluster compactness and distinctiveness.}
    \label{fig:cluster_metrics_comparison}
\end{figure*}

%% Alternative validation metrics
\noindent \textbf{Validation metrics.}
\rev{To further assess the appropriateness of the selected number of clusters, we compared spectral clustering across a range of cluster numbers ($k=2$ to $k=9$) using three internal validation metrics: Silhouette score, Calinski-Harabasz (CH) index, and Davies-Bouldin (DB) index. The Silhouette and CH indices evaluate cluster compactness and separation (higher values indicate better-defined clusters), while the DB index measures average cluster similarity (lower values indicate better separation).
Results are seen on \autoref{fig:cluster_metrics_comparison}. The Silhouette score reaches its maximum at $k=2$ and decreases for larger values of $k$, indicating that additional clusters reduce average separation in Euclidean space. A similar declining pattern is observed for the CH index. The DB index fluctuates across specifications and does not exhibit a pronounced global minimum at higher $k$ - however, its minimum is achieved at $k=3$.
Taken together, the three metrics consistently indicate stronger cluster separation for smaller values of $k$, with $k=3$ representing a reasonable balance between cluster compactness and interpretability. Importantly, while internal validation metrics naturally decline as cluster granularity increases, the overall patterns are consistent, supporting the stability and structural coherence of the identified park classes.}

%------------------------
%------------------------
\section{Image Sharing Varies Among Park Types}
\label{subsubsec:clusters_flickr}
\begin{table*}[t]
\caption{Relative importance of Flickr tags within the three identified groups of parks. The values after the tag indicate how much more frequent these tags were compared to a uniform distribution (Section 4.8 of the main text). }
\label{tab:park_tags}
\centering
\begin{tabularx}{\textwidth}{rXXX}
  \hline
 & \textbf{Lunchbreak Parks} & \textbf{Cultural Parks} & \textbf{Recreational Parks} \\ 
  \hline
  
   % &\includegraphics[width=\linewidth]{figures/nature/lunchbreak_park.jpg}&\includegraphics[width=\linewidth]{figures/nature/cultural_park.jpg}&\includegraphics[width=\linewidth]{figures/nature/recreational_park.jpg}\\
   % &Parc André Citroën, \small CC-BY Guilhem Vellut & Champ de Mars, \small  CC-BY Jacobo Tarrío & Parc Georges-Valbon, \small CC-BY Guilhem Vellut \\
   % \hline
1 & plant, 2.73 & bercy, 3.72 & beach, 3.84 \\ 
  2 & arbre, 2.19 & music, 2.32 & arbre, 3.66 \\ 
  3 & architecture, 2.13 & sculpture, 1.82 & iledefrance, 3.17 \\ 
  4 & text, 2.01 & festival, 1.68 & automne, 2.57 \\ 
  5 & face, 2.01 & live, 1.66 & art, 1.83 \\ 
  6 & tree, 2 & concert, 1.61 & spring, 1.73 \\ 
  7 & water, 1.83 & iphoneography, 1.51 & tree, 1.69 \\ 
  8 & film, 1.79 & instagramapp, 1.51 & fleurs, 1.59 \\ 
  9 & jardin, 1.72 & tree, 1.48 & nature, 1.59 \\ 
  10 & garden, 1.7 & fleurs, 1.44 & fleur, 1.27 \\
   & automne, 1.66; people, 1.61; iphone, 1.55; bercy, 1.48; outdoor, 1.46; blue, 1.41; fleurs, 1.39; urban, 1.39; iledefrance, 1.34; autumn, 1.32; instagramapp, 1.31; iphoneography, 1.3; beach, 1.29; landscape, 1.29; mountain, 1.25; flower, 1.24; trees, 1.22; street, 1.17; statue, 1.15; geotagged, 1.14; nature, 1.13; sculpture, 1.12; car, 1.1; museum, 1.09; flowers, 1.08; winter, 1.07; nuit, 1.07; spring, 1.03; white,~1.01 & text, 1.37; automne, 1.35; seine, 1.27; face, 1.25; geotagged, 1.22; architecture, 1.21; art, 1.2; arbre, 1.2; iphone, 1.19; street, 1.18; water, 1.16; iledefrance, 1.09; plant, 1.09; outdoor, 1.08; spring, 1.06 & \\
   \hline
\end{tabularx}
\end{table*}

To provide an alternative perspective and assess whether shared images support the functional roles of parks identified through traffic data, we analyzed geo-located images from Flickr. These images were collected via the Flickr API in 2015 and include $539,763$ photos across the parks in our study, an average of $12,851$ images per park. Each image could be tagged with an arbitrary number of labels, both user-generated and algorithmically assigned\cite{Li2009Towards,Thomee2016YFCC100M}. On average, each image has $7.45$ tags, resulting in a total of $4,022,720$ tags, approximately $95,779$ per park.
To measure the relative importance of each tag, we calculated the ratio between its actual and expected frequency. Expected frequencies were estimated using a uniform distribution of tag occurrences across parks (see Section 4.8 of the main text for details). A relative importance value of 1 indicates an average occurrence, while values above $1$ suggest the tag appears more often than expected.
The relative importance of tags for all parks in each cluster is shown in \autoref{tab:park_tags}. Tags are listed in descending order of importance within each cluster, with a cutoff value of 1. The values following each tag represent the mean relative importance in that cluster.

\textbf{Lunchbreak parks} show the highest number of tags occurring above expectation (39), highlighting a diverse image set. Tags relate both to natural elements (e.g., `plant', `arbre'—French for tree, `garden') and to urban features, such as `architecture', `text' (indicating signage), and `face', which signals the presence of people in photos.

\textbf{Cultural parks} are characterized by tags associated with arts, concerts, and festivals. The presence of `instagramapp’ suggests that many images were also shared on Instagram. Overall, the tag patterns reinforce the cultural and touristic appeal of these parks.

\textbf{Recreational parks} have the fewest tags above expectation ($n=10$), reflecting lower tag diversity. Many of these tags are in French, suggesting that the user-contributed tags come from a predominantly local user base.

\noindent\textbf{Predictive Power of Flickr Tags.}
To quantitatively assess the relevance of tag frequency, we conducted a prediction experiment to determine whether Flickr tags can be used to classify the functional uses of parks.
We implemented a $3$-fold cross-validated Random Forest classifier with $100$ estimators, using a feature vector representing the relative importance of the 25 most frequent tags per park. The prediction target was the park’s assigned cluster. Our model achieved accuracies ranging between $[0.571; 0.715]$, with a mean accuracy of~$0.643$ and an $F_1$ score of $0.619$.
Our model performs significantly better than chance ($0.333$), confirming that images shared in parks with different functional uses exhibit distinct patterns. However, its predictive power is only moderate. This is likely because Flickr tags alone are insufficient for reliably classifying park functions, due to the noisy nature of tagging data and imbalances in the number of photos per park. Additionally, there is a temporal gap of approximately 10 years between the two datasets.

% --
\section{Supplementary Tables}
% --

% % --
% \section{Statistical tests}
% \label{subsec:stat_tests}
% % --

We next describe additional statistical tests performed for the insights gained throughout Section 2 of the main text. 
The first test seeks to understand if the distribution of RSCA values across application categories differs within all parks in Paris, versus the remaining areas of the city. We first use a Levene's test to assess if the distributions' variances are equal. With this result, we select a $t$-test to validate if the distributions are statistically different. A Levene $p$-value below $0.05$ leads to a Welch's $t$-test as the distributions have unequal variances, while a $p$-value above $0.05$ leads to a Student's $t$-test (as we can assume the distributions have equal variances).
Through the $t$-tests, we find that 4 application categories have a significant statistical difference in their preferences in parks versus the rest of the city: Games and News have $p<0.05$ and Music and Social have $p<0.01$. All results are detailed in \autoref{tab:significant_dif_tests}.

\begin{table*}[h]
    \centering
    \caption{
    T-test results assessing significant differences in per-category RSCA between parks and the city. 
    Bold values represent that there is a significant difference between park and city, while asterisks represent the intensity  (* for $p<0.05$, ** for $p<0.01$).
    A Levene's test was first applied to check variance similarity, in order to select the correct t-test.
    }
    \begin{tabular}{@{}lccc|cc@{}}
        \toprule
        \textbf{Application Category} & \textbf{Levene $p$-value} & \textbf{Selected t-test} & & \textbf{$p$-value} \\ 
        \midrule
        Fitness & 0.104 & Student's & & 0.214  \\
        Games & 0.025 & Welch’s & & \textbf{0.027*} \\
        Music & 0.907 & Student's  & & \textbf{0.005**} \\
        News and Info & 0.344 & Student's & & \textbf{0.049*} \\
        Productivity & 0.549 & Student's & & 0.112  \\
        Shopping & 0.004 & Welch’s & & 0.086  \\
        Social & 0.491 & Student's & & \textbf{0.005**} \\
        Travel & 0.194 & Student's & & 0.164  \\
        Video & 0.840 & Student's & & 0.077  \\ \bottomrule
    \end{tabular}
    \label{tab:significant_dif_tests}
\end{table*}

Next, we replicate the previous experiment but with park clusters, to understand if each cluster has statistically significant application preference versus the rest of the city. The full results are presented in \autoref{tab:significant_dif_tests2}.%, with the details of the statistical significant cases discussed on \autoref{subsec:clusters_apps}. 

\begin{table}[ht]
    \centering
    \caption{T-test results assessing significant differences in per-category RSCA between parks and the city. 
    Bold values represent that there is a significant difference between park and city, while asterisks represent the intensity  (* for $p<0.05$, ** for $p<0.01$).
    A Levene's test was first applied to check variance similarity, in order to choose the correct t-test.}
    \begin{tabular}{@{}lccc|ccc@{}}
        \toprule
        \textbf{Application Category} & \textbf{Cluster} & \textbf{Levene p-value} & \textbf{Selected t-test} & \textbf{t-test p-value} \\
        \midrule
        Fitness & Recreational & 0.820 & Student & \textbf{0.029*} \\
        & Cultural & 0.010 & Welch & 0.711 \\
        & Lunchbreak & 0.347 & Student & 0.351 \\
        \midrule
        Games & Recreational & 0.602 & Student & 0.052 \\
        & Cultural & 0.016 & Welch & \textbf{0.048*} \\
        & Lunchbreak & 0.256 & Student & \textbf{0.040*} \\
        \midrule
        Music & Recreational & 0.314 & Student & \textbf{0.004**} \\
        & Cultural & 0.005 & Welch & \textbf{$<$0.001**} \\
        & Lunchbreak & 0.044 & Welch & \textbf{0.007**} \\
        \midrule
        News & Recreational & 0.284 & Student & \textbf{0.017*} \\
        & Cultural & 0.011 & Welch & 0.956 \\
        & Lunchbreak & 0.496 & Student & 0.929 \\
        \midrule
        Productivity & Recreational & 0.003 & Welch & \textbf{0.013*} \\
        & Cultural & $<$0.001 & Welch & 0.063 \\
        & Lunchbreak & 0.009 & Welch & 0.149 \\
        \midrule
        Shopping & Recreational & 0.537 & Student & \textbf{0.020*} \\
        & Cultural & 0.481 & Student & 0.656 \\
        & Lunchbreak & 0.053 & Student & 0.196 \\
        \midrule
        Social & Recreational & 0.678 & Student & \textbf{$<$0.001**} \\
        & Cultural & 0.683 & Student & 0.856 \\
        & Lunchbreak & 0.145 & Student & 0.342 \\
        \midrule
        Travel & Recreational & 0.440 & Student & \textbf{$<$0.001**} \\
        & Cultural & 0.002 & Welch & 0.054 \\
        & Lunchbreak & 0.285 & Student & 0.099 \\
        \midrule
        Video & Recreational & 0.769 & Student & \textbf{$<$0.001**} \\
        & Cultural & 0.067 & Student & \textbf{0.009**} \\
        & Lunchbreak & 0.309 & Student & 0.368 \\
        \bottomrule
    \end{tabular}
    \label{tab:significant_dif_tests2}
\end{table}

Finally, we explore for each park cluster if there is a significant correlation between application preference and socioeconomic indicators. We calculate the Pearson correlation between values, and consider a correlation statistically relevant if p$<$0.05. Results are fully detailed in \autoref{tab:correlation_rsca}, with results discussed in Section 2.4 of the main text.

\begin{table}[ht]
    \centering
    \caption{Correlation statistics and p-values for socioeconomic indicators across park categories and selected application categories. Bold values indicate statistically significant correlations ($p \leq 0.05$).
    \label{tab:correlation_rsca}
    }
    % Median Income
    \begin{tabular}{lccc}
        \hline
        \textbf{} & \multicolumn{3}{c}{\textbf{Median Income}}\\
        \hline
        \textbf{Indicator} & \textbf{Cultural parks} & \textbf{Lunchbreak parks} & \textbf{Recreational parks} \\
        \hline
        Games & 0.093 (0.723) & -0.273 (0.416) & \textbf{-0.672 (0.003)} \\
        Music & 0.322 (0.208) & \textbf{0.728 (0.011)} & \textbf{0.608 (0.010)} \\
        News and Information & \textbf{0.606 (0.010)} & -0.092 (0.788) & 0.317 (0.215) \\
        Social & -0.188 (0.471) & 0.254 (0.450) & -0.067 (0.798) \\
        \hline
        \vspace{0.5mm}
    \end{tabular}
    % Gini Index
    \begin{tabular}{lccc}
        \hline
        \textbf{} & \multicolumn{3}{c}{\textbf{Gini Index}}\\
        \hline
        \textbf{Indicator} & \textbf{Cultural parks} & \textbf{Lunchbreak parks} & \textbf{Recreational parks} \\
        \hline
        Games & 0.153 (0.559) & -0.259 (0.442) & 0.264 (0.306) \\
        Music & 0.295 (0.250) & \textbf{0.691 (0.018)} & -0.035 (0.893) \\
        News and Information & \textbf{0.697 (0.002)} & 0.060 (0.861) & 0.213 (0.412) \\
        Social & -0.007 (0.979) & 0.293 (0.383) & -0.212 (0.414) \\
        \hline
        \vspace{0.5mm}
    \end{tabular}
    % Distance 
    \begin{tabular}{lccc}
        \hline
        \textbf{} & \multicolumn{3}{c}{\textbf{Distance}}\\
        \hline
        \textbf{Indicator} & \textbf{Cultural parks} & \textbf{Lunchbreak parks} & \textbf{Recreational parks} \\
        \hline
        Games & -0.063 (0.810) & 0.392 (0.233) & 0.379 (0.134) \\
        Music & -0.449 (0.070) & -0.318 (0.340) & -0.394 (0.118) \\
        News and Information & -0.212 (0.415) & 0.086 (0.801) & -0.234 (0.365) \\
        Social & -0.121 (0.645) & -0.560 (0.073) & -0.038 (0.884) \\
        \hline
        \vspace{0.5mm}
    \end{tabular}
    % Park size 
    \begin{tabular}{lccc}
        \hline
        \textbf{} & \multicolumn{3}{c}{\textbf{Park size}}\\
        \hline
        \textbf{Indicator} & \textbf{Cultural parks} & \textbf{Lunchbreak parks} & \textbf{Recreational parks} \\
        \hline
        Games & \textbf{0.589 (0.013)} & 0.487 (0.129) & \textbf{0.608 (0.010)} \\
        Music & 0.466 (0.059) & 0.091 (0.790) & -0.095 (0.716) \\
        News \& Information & 0.147  (0.574) & -0.070 (0.839) & 0.326 (0.202) \\
        Social & -0.317 (0.216) & -0.555 (0.076) & -0.063 (0.811) \\
        \hline    
    \end{tabular}
\end{table}